\documentclass[%
superscriptaddress,
preprint,
onecolumn,
nofootinbib,
 amsmath,amssymb,
 aps,
]{revtex4-2}
\usepackage{color}
\usepackage{graphicx}% Include figure files
\usepackage{dcolumn}% Align table columns on decimal point
\usepackage{bm}% bold math
\usepackage{comment}
\begin{document}

%\preprint{???}

\title{Localization of Electronic States in Hybrid Nano-Ribbons in the Non-Perturbative Regime }% Force line breaks with \\

\author{Thomas Luu}%
\email{t.luu@fz-juelich.de}
\affiliation{Institute for Advanced Simulation (IAS-4),
and J\"ulich Center for Hadron Physics, Forschungszentrum J\"ulich, Germany}
%\affiliation{Helmholtz-Institut f\"ur Strahlen- und Kernphysik %and Bethe Center %for Theoretical Physics, Rheinische Friedrich-Williams-Universit\"at Bonn, %Germany}%

\author{Ulf-G. Mei{\ss}ner}
\email{meissner@hiskp.uni-bonn.de}
\affiliation{Helmholtz-Institut f\"ur Strahlen- und Kernphysik and Bethe Center for Theoretical Physics, Rheinische Friedrich-Williams-Universit\"at Bonn, Germany}
\affiliation{Institute for Advanced Simulation (IAS-4),
and J\"ulich Center for Hadron Physics, Forschungszentrum J\"ulich, Germany}
%\affiliation{Center for Science and Thought, Rheinische %Friedrich-Williams-Universit\"at Bonn, Germany}
\affiliation{Tbilisi State University, 0186 Tbilisi, Georgia}

\author{Lado Razmadze}
\email{s6larazm@uni-bonn.de}
\affiliation{Helmholtz-Institut f\"ur Strahlen- und Kernphysik and Bethe Center for Theoretical Physics, Rheinische Friedrich-Williams-Universit\"at Bonn, Germany}%

\date{\today}% It is always \today, today,
\begin{abstract}
We investigate the localization of low-energy single quasi-particle states in the 7/9-hybrid nanoribbon system in the presence of strong interactions and within a finite volume.  We consider two scenarios, the first being the Hubbard model at half-filling and perform quantum Monte Carlo simulations for a range $U$ that includes the strongly correlated regime.   In the second case we add a nearest-neighbor superconducting pairing $\Delta$ and take the symmetric line limit, where $\Delta$ is equal in magnitude to the hopping parameter $t$. In this limit the quasi-particle spectrum and wavefunctions can be directly solved for general onsite interaction $U$. In both cases we extract the site-dependent quasi-particle wavefunction densities and demonstrate that localization persists in these non-perturbative regimes under particular scenarios.
\end{abstract}

\maketitle
\pagebreak
%\tableofcontents

\section{Introduction\label{sec:intro}}
Recently it was shown that localized, low-energy states can occur at the junction of two nanoribbons that are topologically distinct~\cite{PhysRevLett.119.076401}.  The presence of such symmetry-protected topological (SPT) localized states depends on their junction geometry and topological invariance. The ability to engineer such hybrid ribbons~\cite{rizzo18,groening2018} has spurred research into the use of these systems for manufacturing quantum dots~\cite{doi:10.1021/acsnano.1c09503}, potentially leading the way to novel, advanced electronic devices and an avenue for obtaining fault-tolerant quantum computing.

The presence of these localized states is manifest in the non-interacting, tight-binding scenario. Furthermore, SPT protection is only strictly enforced when both ribbons extend infinitely from the junction, as the topological invariants are calculated for infinite armchair graphene nanoribbons (AGNRs). The system has a very small energy gap compared to the hopping parameter. Though~\cite{PhysRevLett.119.076401} have demonstrated the stability of such states under perturbation, the extent to which these SPT states remain low energy, as well as localized, in the strongly interacting regime is an open question, especially since any practical implementation of these hybrid systems will be finite in extent, or perhaps in a repeating lattice.

In this paper we address the question of finite volume directly by investigating a \emph{periodic} 7/9-hybrid nanoribbon in two distinct non-perturbative regimes.  The first considers the standard Hubble model applied to this system at the electrically neutral, half-filling case.  Here we perform quantum Monte Carlo (QMC) simulations for various values of the onsite coupling $U$  that include the strongly interacting regime.     In the second case  we consider  the so-called symmetric line limit~\cite{Yang:2020lal,Ezawa2018,Miao:2019tng}, where we introduce a nearest neighbor superconducting pairing term $\Delta$ to the Hubbard model but with equal weight as the hopping term $t$.  In this limit the single-particle spectrum and wavefunctions, when expressed in a Majorana basis, can be determined for any value of the Hubbard onsite interaction $U$. In both cases we observe that the energy of the localized state is no longer depends strongly on the coupling $U$.  However, it still remains the lowest energy state of the system. Further, we find that under certain conditions the localization of these states at the junction persists.

Our paper is organized as follows.  In the Sect.~\ref {sect:geometry} we describe our 7/9 hybrid lattice geometry and provide solutions to its spectrum in the tight-binding, or non-interacting, limit.  We then introduce a Hubbard onsite interaction $U$ in Sect.~\ref{sect:qmc} and show results of our QMC simulations for select values of $U$.  In Sect.~\ref{sect:symLine} we consider the symmetric line limit by adding a nearest neighbor superconducting term $\Delta$ of equal magnitude to the hopping term and apply it to this particular 7/9 hybrid nanoribbon.  We demonstrate how this system can be solved directly for any value of $U$ and show the dependence of the energy and wavefunction of the localized state on $U$.    These localized states on opposing sublattices, or chiralities, have a potential connection to domain-wall fermions formulated in lattice gauge theories in 4+1 dimensions~\cite{Kaplan:1992bt,Shamir:1993zy}.  We comment on this potential connection in Sect.~\ref{sect:dmf}.  We recapitulate in Sect.~\ref{sect:conclusions}.

\section{Geometry of the periodic 7/9 hybrid nanoribbon\label{sect:geometry}}

Unit cells in such AGNRs are defined by their terminations i.e. shapes of their edges. In \cite{PhysRevLett.119.076401} four distinct types of unit cells were defined. Based on inversion and mirror symmetries, as well as the width of ribbons, it was been shown that such systems have an associated conserved quantity, the so-called $Z_2$ topological invariant, that can take  the values 0 or 1. The interface of two materials with distinct topological invariants can support surface modes \cite{PhysRevB.95.035421}. Since the existence of these modes depend solely on the topological factors, they should remain even under the presence of interactions, given that these interactions do not change the invariants themselves. The main example used in this paper is the 7/9 hybrid nanoribbon, where part of the ribbon with width 7 has topological invariant $Z_2=0$, while the part with width 9 has invariant $Z_2=1$.

\begin{figure}
    \centering
    \includegraphics[width=\textwidth]{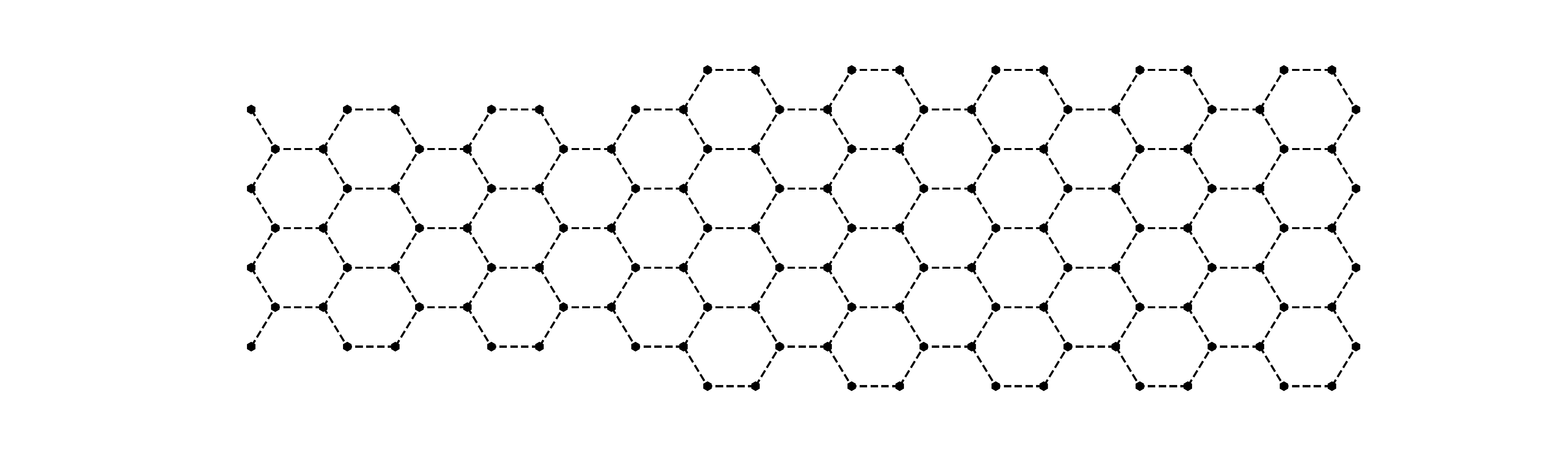}
    \caption{Single unit cell of the 7/9 hybrid system considered in this work.  The widths are set by the 7 and 9 armchair nanoribbon parts, while the lengths have 6 hexagons and 10 hexagons for the 7 and 9 parts, respectively.  Periodic boundary conditions are employed at the ends.}
    \label{fig:geometries}
\end{figure}

This 7/9 hybrid system is shown in Fig.~\ref{fig:geometries}.  This system, representing a single unit cell, has $N=132$ total lattice sites and is composed of six hexagonal units lengthwise for the 7 AGNR part, and 10 hexagons lengthwise for the 9 AGNR part.  In terms of the lattice spacing $a$ between sites, the entire length of the unit cell is $L=24a$.  The system is bipartite, meaning we can divide the lattice into two independent sublattices, which we label one as consisting of A sites, and the other B sites.  We apply periodic boundary conditions at the ends so that the unit cell shown in Fig.~\ref{fig:geometries} repeats itself.  

\begin{figure}
    \centering
    \includegraphics[width=.8\textwidth]{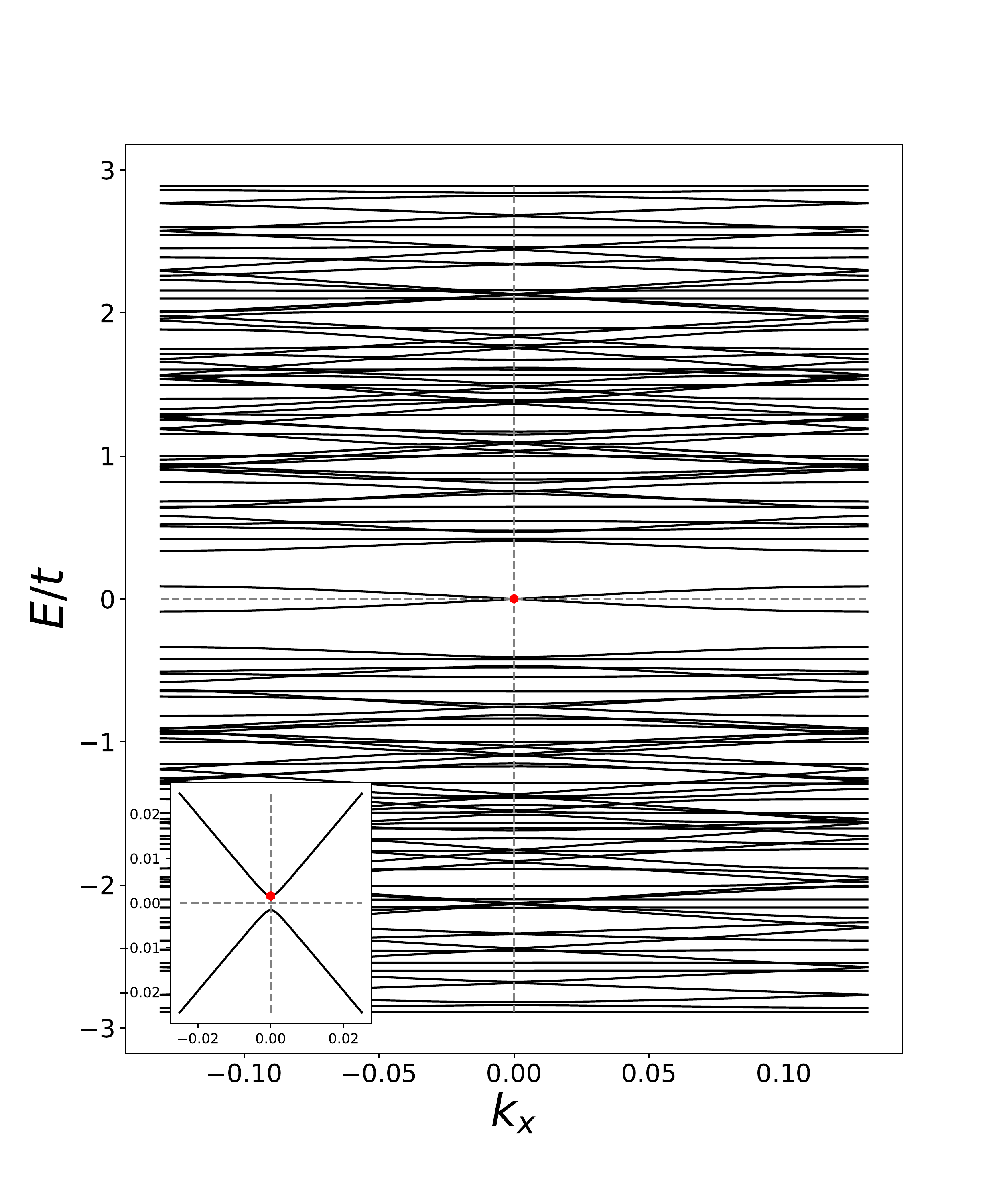}
    \caption{Non-interacting dispersion of the 7/9 hybrid ribbon. The inset shows the avoided level crossing near the Fermi surface at $k_x=0$.  $k_x$ is expressed in units of the inverse length $L^-1$ with $L=24a$ being the unit cell length and $a$ the lattice spacing. \label{fig:7_3.9_5 dispersion}}
\end{figure}

\begin{figure}
    \centering
    \includegraphics[width=\textwidth]{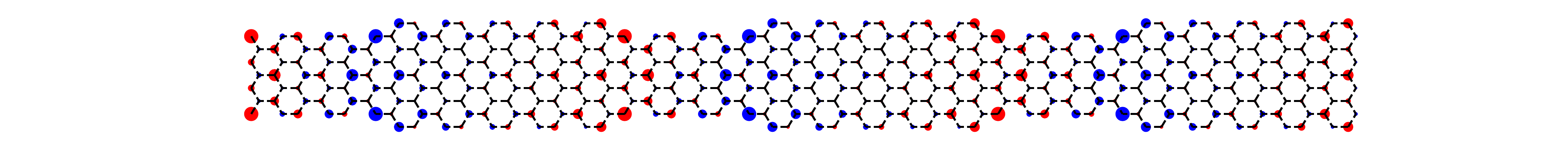}
    \caption{Single-particle wavefunction densities for the state with energies $E/t = 0.0015996$  in the non-interacting limit.  The size of the circles is proportional to the density and the color denotes the two sublattices, red = A sites, blue = B sites.}
    \label{fig:79AGNR}
\end{figure}

Under the tight-binding approximation, or equivalently the non-interacting limit, we have
\begin{equation}\label{eqn:H0}
H_0=-t\sum_{\langle i,j\rangle,\sigma} a^\dag_{i\sigma}a^{}_{j\sigma}+h.c.\ ,
\end{equation}
where $t$ is the hopping parameter, $a^\dag_j$ ($a^{}_j$) is the fermionic creation (annihilation) operator at lattice site $j$, $\sigma$ the spin, $h.c.$ stands for Hermitian conjugate, and the sum is over all nearest neighbors $\langle i,j\rangle$.  As the Hamiltonian is quadratric in the number of creation and annihilation operators, the single particle dispersion as a function of longitudinal momentum $k_x$ can be easily determined in this limit, which we show in Fig.~\ref{fig:7_3.9_5 dispersion}.  Note that the dispersion is the same for both spins.
Of particular interest is the point at $k_x=0$ where there seems to be an apparent level crossing at $E/t=0$, denoted as a red point in the main plot in Fig.~\ref{fig:7_3.9_5 dispersion}.   In fact, upon closer inspection as shown in the inset of Fig.~\ref{fig:7_3.9_5 dispersion}, there is \emph{no} level crossing at this point, as shown in the inset, since it consists of two states with energies
\begin{equation}
    E/t = \pm 0.0015996\ .%55851370325\ .
\end{equation}
These states exhibit localization at the junctions of the 7 and 9 AGNR parts.  In Fig.~\ref{fig:79AGNR} we show the wavefunction densities, $\rho(x)=|\psi(x)|^2$, for each lattice site on an extended hybrid system for this particular state.  These densities are the same for either positive or negative energy solutions.  The localization of the states at the junctions is apparent in this figure.
Further, the localization is confined to specific sublattices denoted by the red and blue colors in Fig.~\ref{fig:79AGNR}, and alternates between the different junctions.  

In what follows, we assume that the system is electrically neutral and thus half-filled,  meaning that all negative energy states are occupied.  The Fermi surface of the system then corresponds to zero energy. We thus concentrate on the lowest \emph{unoccupied} single-particle state.  In the non-interacting case this corresponds to the state denoted by the red dot in the inset of Fig.~\ref{fig:7_3.9_5 dispersion}.  In the remainder of the paper we loosely refer to this state as the localized state, though it remains to be seen if the state remains localized in the presence of interactions and within a finite volume.

\section{QMC calculations of the Hubbard Model\label{sect:qmc}}
We now include a Hubbard onsite interaction,
\begin{equation}
    H_=-\sum_{\langle i,j\rangle,\sigma}\left(t_{ij} a^\dag_{i\sigma}a^{}_{j\sigma}+h.c.\right)+U\sum_x\left(n_{x\uparrow}-\frac{1}{2}\right)\left(n_{x\downarrow}-\frac{1}{2}\right)\ .
\end{equation}
The form of the interaction ensures that the system remains at half-filling.  Note that the onsite interaction is quartic in the number of creation and annihilation operators, and therefore no direct diagonalizaton is possible.  Therefore we use QMC simulations to investigate the hybrid system for values of the onsite interaction $U$ corresponding to the strongly coupled regime.  Our formalism for performing QMC simulations of low-dimensional Hubbard systems have been described in detail in \cite{Luu:2015gpl,Ostmeyer:2020uov,Ostmeyer:2021efs}.  Here we just point out some salient features pertinent to this work.  

\begin{figure}
    \centering
    \includegraphics[width=.8\textwidth]{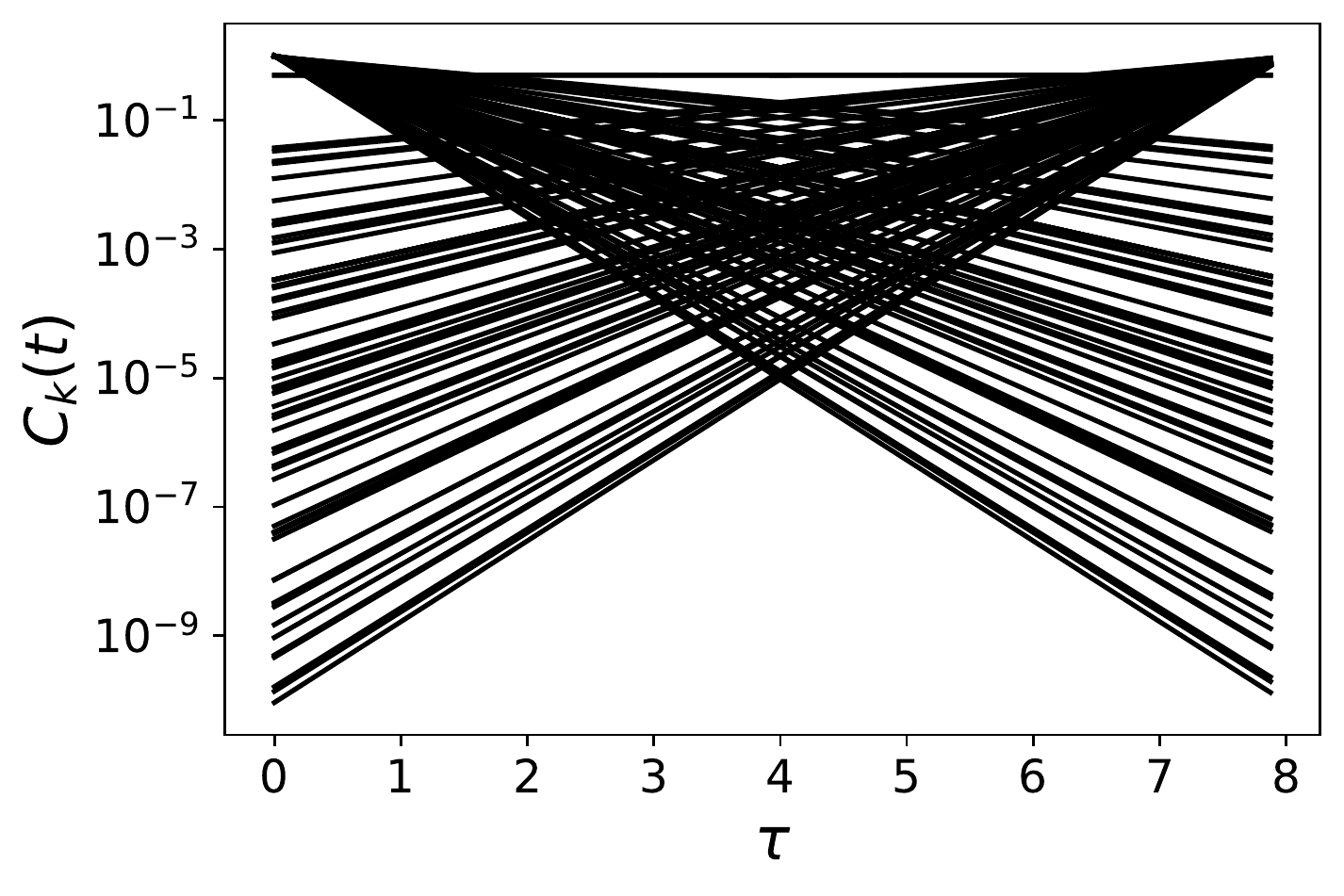}
    \caption{All non-interacting correlators at $k_x=0$.\label{fig:corr_ni}}
\end{figure}

To extract the quasi-particle energies we calculate momentum correlators as a function of time,
\begin{equation}\label{eqn:Ck}
    C_k(\tau)\equiv \langle a^{}_k(\tau)a^\dag_k(0)\rangle=\frac{1}{Z} \operatorname{Tr}\ \left[a_k^{}(\tau)a^\dag_k(0)e^{-\beta H}\right]\ ,
\end{equation}
where $\beta$ represents an inverse temperature and $k=(k_x,\kappa)$ is a momentum index corresponding to the state. The time $\tau\in[0,\beta]$, and in our simulations we discretize this variable into $N_t$ timeslices.  We use $N_t=64,\ 80$ and $96$ in our simulations with $\beta=8,\ 10,$ and $12$, respectively. The variable $\kappa$ is an index corresponding to one of the possible $N=132$ states of the system for a given $k_x$.  We choose $\kappa$ to correspond to the state with the lowest possible positive energy.  Fig.~\ref{fig:corr_ni} shows all $k_x=0$ correlators at the non-interacting $U=0$ case and $\beta=8$.
A spectral decomposition of the expression in Eq.~\eqref{eqn:Ck}, as is done in App.~\ref{sect:densities}, shows that these correlators have an exponential dependence in time, $\sim e^{-E_k \tau}$, where their arguments correspond to the non-interacting energies of the system at $k_x=0$.  These energies correspond to both the positive and negative points that occur at $k_x=0$ in Fig.~\ref{fig:7_3.9_5 dispersion}. 

In the presence of interactions $U\ne 0$, and within a finite inverse temperature $\beta$, the correlators will have a more complicated dependence on $\tau$ due to thermal contamination with excited states and backwards-propagating states.  However, these effects are usually fleeting since the dependence on the excited states is exponentially suppressed.   Therefore, as long as $\tau\gg 1$ but $\tau<\beta$, the correlators will recover an exponential dependence but now with fully interacting energies in their arguments.  
\begin{figure}
    \centering
    \includegraphics[width=.5\textwidth]{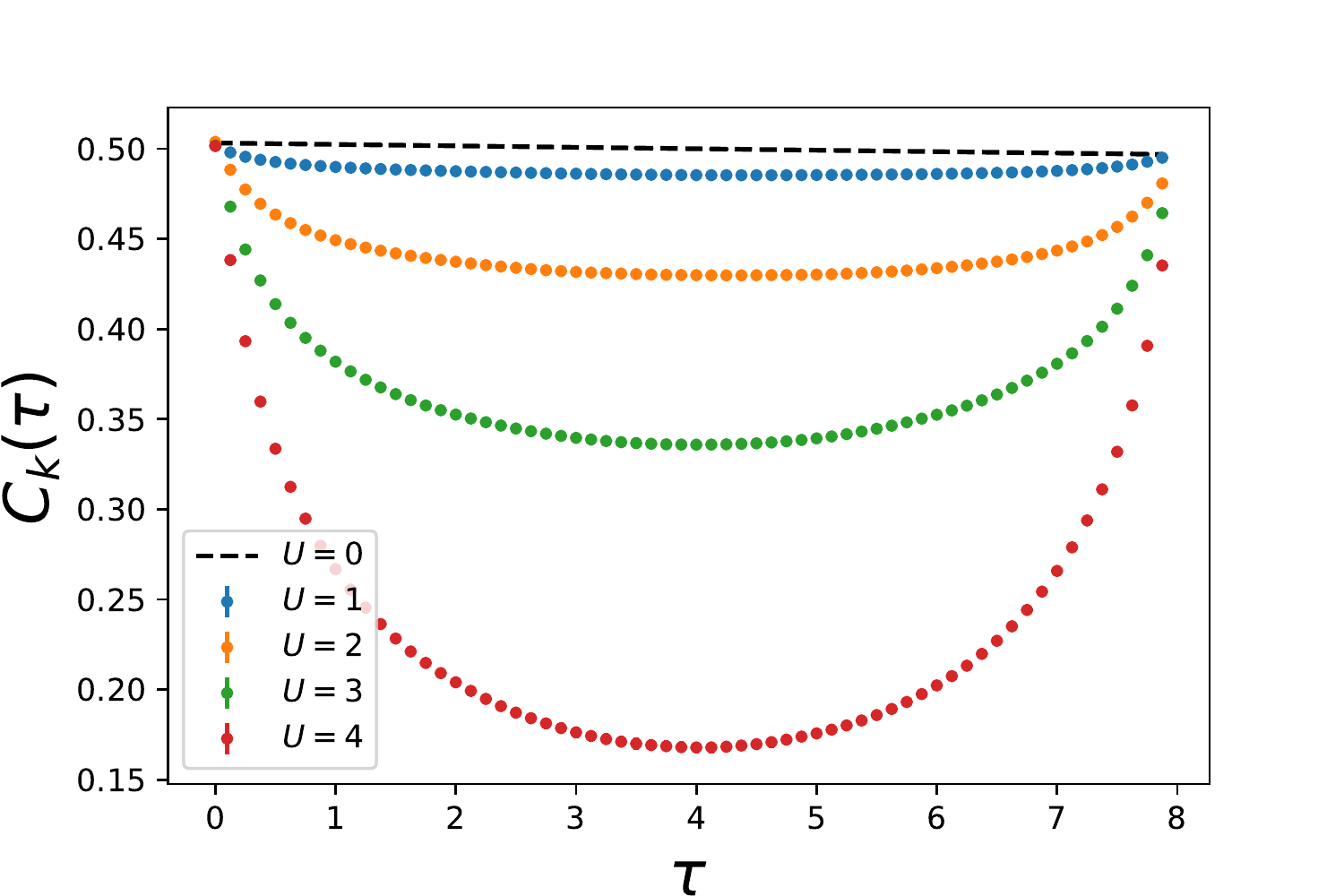}\includegraphics[width=.5\textwidth]{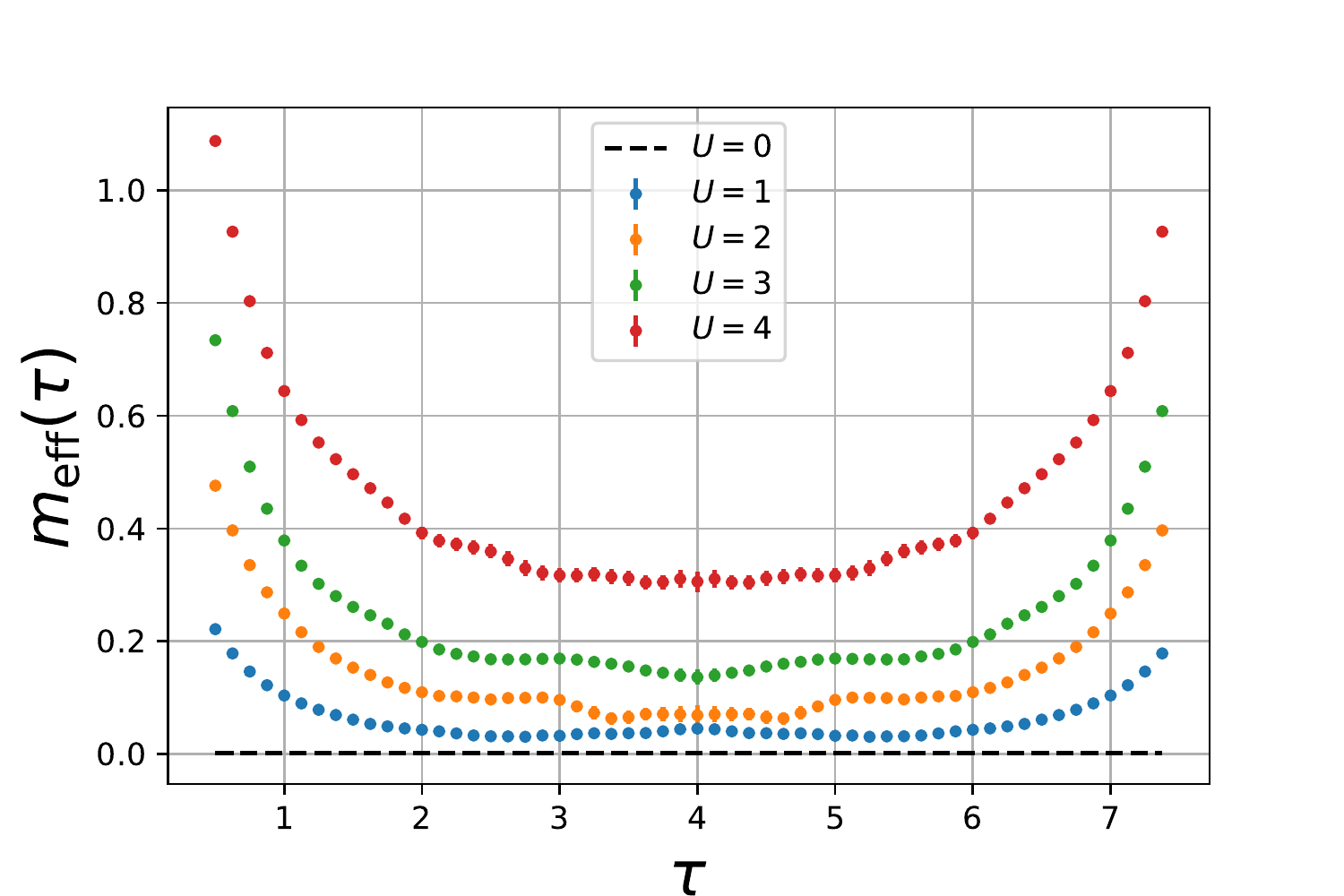}
    \caption{Dependence of correlators for the lowest energy localized state (left) and its corresponding effective masses (right) as defined in Eq.~\eqref{eqn:effmass}.  In both cases the black dashed line corresponds to the non-interacting result.}
    \label{fig:corrU}
\end{figure}
One can thus extract the fully interacting energies by analyzing the exponential behavior of the correlators in this regime.

However, special care must be taken when dealing with correlators that represent states with very small energies, $E\beta\ll 1$, as is the case in our situation.  Here the backwards-propagating states can provide a nearly equally important contribution to the correlator, essentially making the correlator symmetric about the $\tau=\beta/2$ point.  The left panel of Fig.~\ref{fig:corrU} shows examples of the dependence of the correlators for the localized state as a function of $U$.  It is indeed the case that these low-energy correlators cannot be described by a single exponential.  We now describe how we extract energies from these correlators.

\subsection{Energy of the localized states}
To extract the energies from these correlators, we first take advantage of the particle-hole symmetry of our problem that states that for any energy solution $E_k$, there is a corresponding solution with opposite sign, $-E_k$.  This is also evident from our correlators, where for each correlator that falls off in time as $C^-_k(\tau)\sim e^{-E_k\tau}$, there is a growing correlator, corresponding to the energy with opposite sign, of the form $C^+_k(\tau)\sim e^{E_k(\tau-\beta)}$.  We average these two correlators.
\begin{equation}
    C_k^{\rm sym}(\tau)=\frac{1}{2}\left(C^-_k(\tau)+C^+_k(\tau)\right)\ ,
\end{equation}
to effectively make a $\cosh$ function of the form $\cosh\left(E_k(\tau-\beta/2)\right)$ in the region $\tau\gg 1$ and $\tau<\beta$.

As a visual aid to estimating the energies of these correlators, we calculate the so called ``effective mass" $m_{\rm eff}(\tau)$,
\begin{equation}\label{eqn:effmass}
    m_{\rm eff}(\tau)=\frac{1}{\delta}\cosh^{-1}\left(\frac{C_k^{\rm sym}(\tau-\delta)+C_k^{sym}(\tau+\delta)}{2C_k^{\rm sym}(\tau)}\right)\ ,
\end{equation}
where $\delta$ is some free parameter.  If $C_k^{\rm sym}(\tau)$ were exactly a $\cosh$ function then $m_{\rm eff}(\tau)=E_k$ for all $\tau$.  As the $\cosh$ behavior is only valid for $1\ll t\ll\beta$ we expect that the effective mass to `flatten out' around the region $\tau=\beta/2$.  The right panel of Fig.~\ref{fig:corrU} shows our extracted effective masses for the localized state using $\delta = 4\beta/N_t$.  As expected the region around $\tau=\beta/2$ is flat and corresponds to the interacting energy $E_k/t$. We stress, however, that these effective masses are only used as a visual aid for estimating the energies.
\begin{figure}
    \centering
    \includegraphics[width=.8\textwidth]{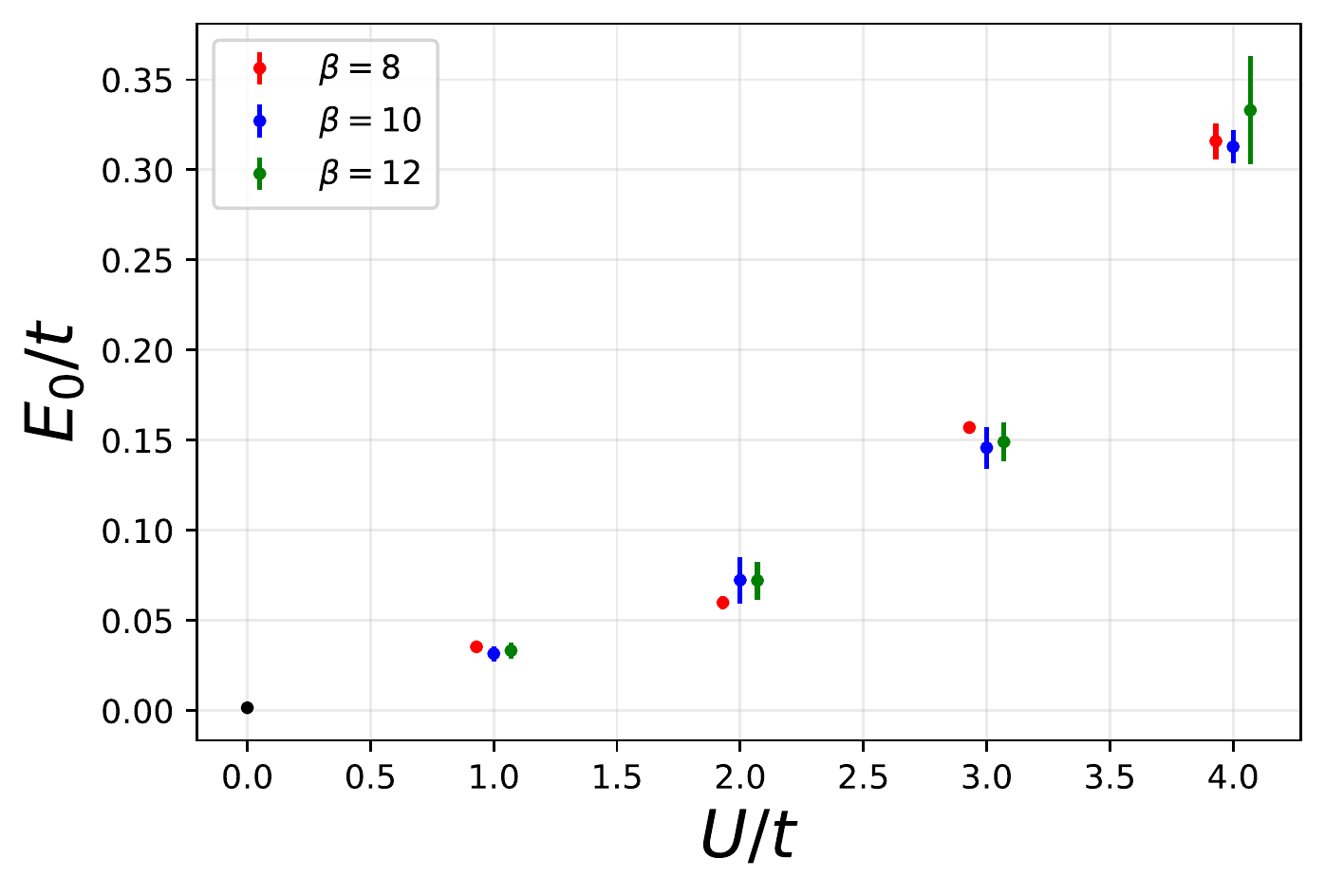}
    \caption{Energy $E_0$ of the lowest state as a function of onsite interaction $U$ obtained from QMC calculations with different values of $\beta$.   The $\beta=8$ (12) results are slightly shifted to the left (right) on the x-axis so as to make the points more easily differentiable.  The black point at $U=0$ is the non-interacting result.  }
    \label{fig:E_vs_U hubbard}
\end{figure}

To actually obtain the energies, we instead we fit directly the correlator $C^{\rm sym}_k(\tau)$.  We show our extracted energies for values of $U\in[1, 2, 3, 4]$ and $\beta\in[8, 10, 12]$ in Fig.~\ref{fig:E_vs_U hubbard}.  Our fits are performed within a finite window around the $\tau=\beta/2$ point and are done under the bootstrap procedure to obtain uncertainties. Looking at Fig.~\ref{fig:E_vs_U hubbard} we see a growing dependence on the energy of the localized state as $U$ increases.  We attribute this dependence to the finite volume of the system, both spatially and temporally, since within such an environment the state is no longer protected by SPT.  Still, in all cases we examined we found that the energy of the localized state remained the lowest, despite its apparent dependence on $U$.  

\subsection{Wavefunction densities of the localized state}

We can also extract the site-dependent densities of the states in our QMC simulations, which in turn allow us to demonstrate localization of the states visually.  A detailed explanation of our calculation is given in App.~\ref{sect:densities}, and we provide only a cursory description here.  

Instead of the momentum correlators calculated in Eq.~\eqref{eqn:Ck}, we instead consider the half-momentum, half-spatial correlators
\begin{equation}\label{eqn:half and half}
   C_k(x,\tau)\equiv \langle a_x^{}(\tau)a_k^\dag(0)\rangle=\frac{1}{Z} \operatorname{Tr}\ \left[a_x^{}(\tau)a^\dag_k(0)e^{-\beta H}\right]\ .
\end{equation}
The effective density $\rho_k(x,\tau)$ for the state $k$ at each lattice site $x$ is given by
\begin{equation}\label{eqn:effective densities}
    \rho_k(x,\tau)\equiv \frac{|C_k(x,\tau)|^2}{\sum_y |C_k(y,\tau)|^2}\ .
\end{equation}
where the sum in the denominator of the right-hand side is over all lattice sites in the unit cell.  As was the case with the effective masses, we extract the densities by looking at the region around $\tau=\beta/2$ where the effective density is flat.  

\begin{figure}
    \centering
    \includegraphics[width=.8\textwidth]{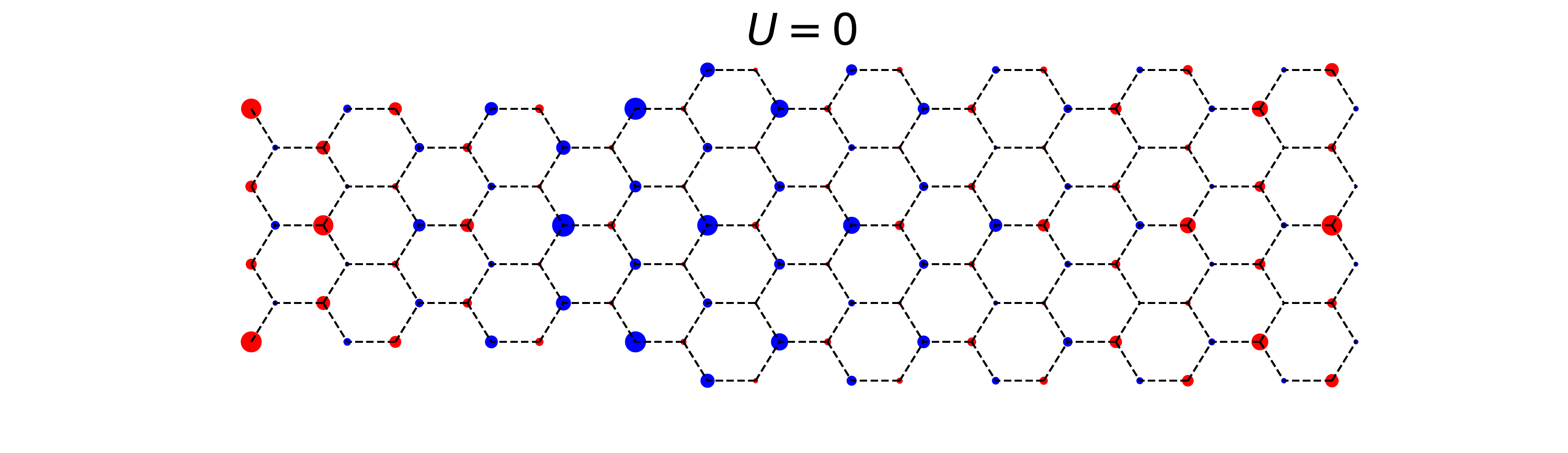}\\
    \includegraphics[width=.8\textwidth]{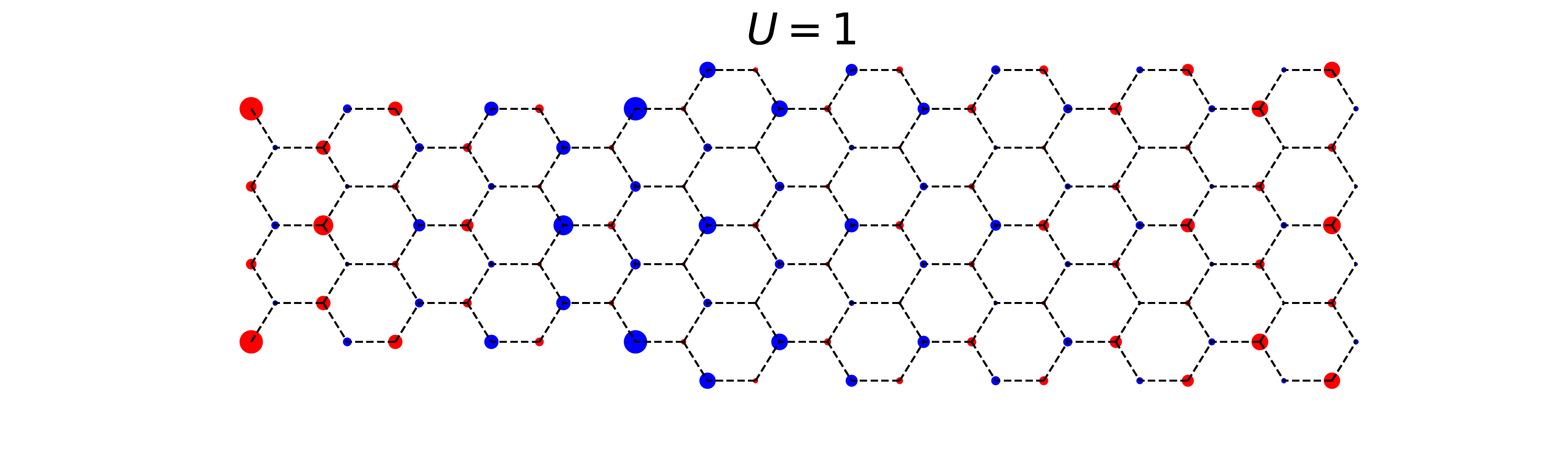}\\
    \includegraphics[width=.8\textwidth]{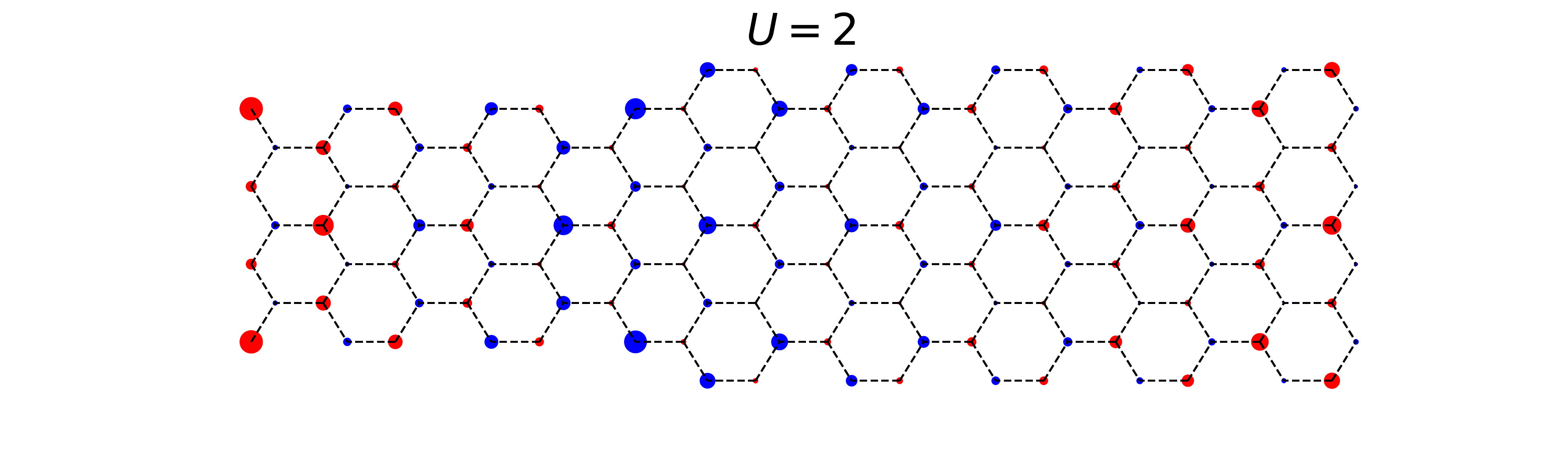}\\
    \includegraphics[width=.8\textwidth]{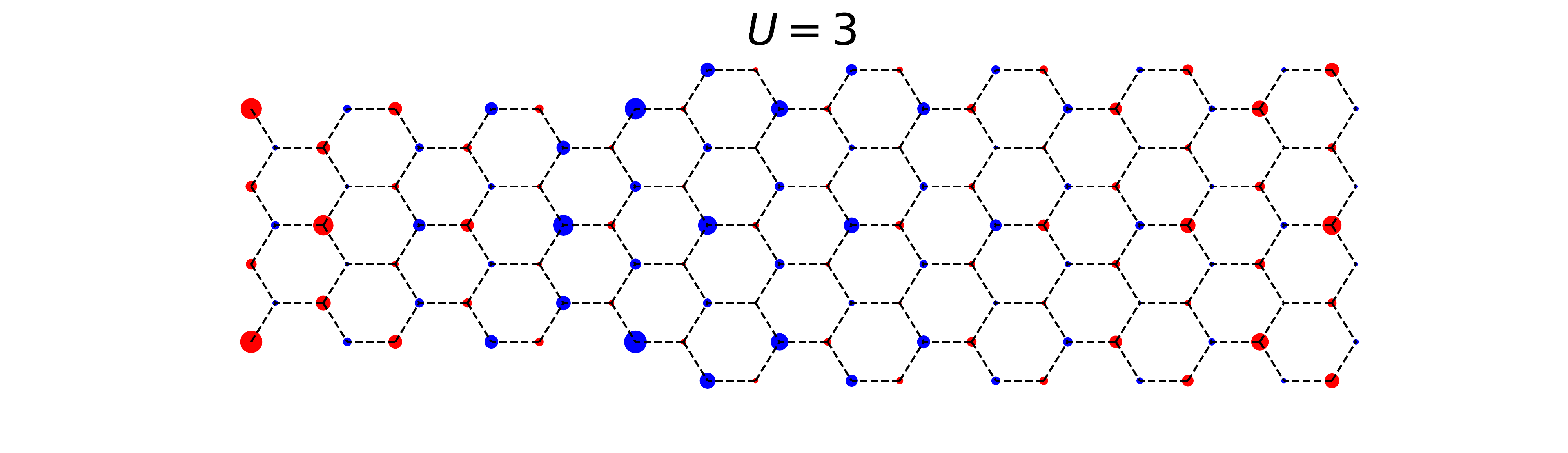}\\
    \includegraphics[width=.8\textwidth]{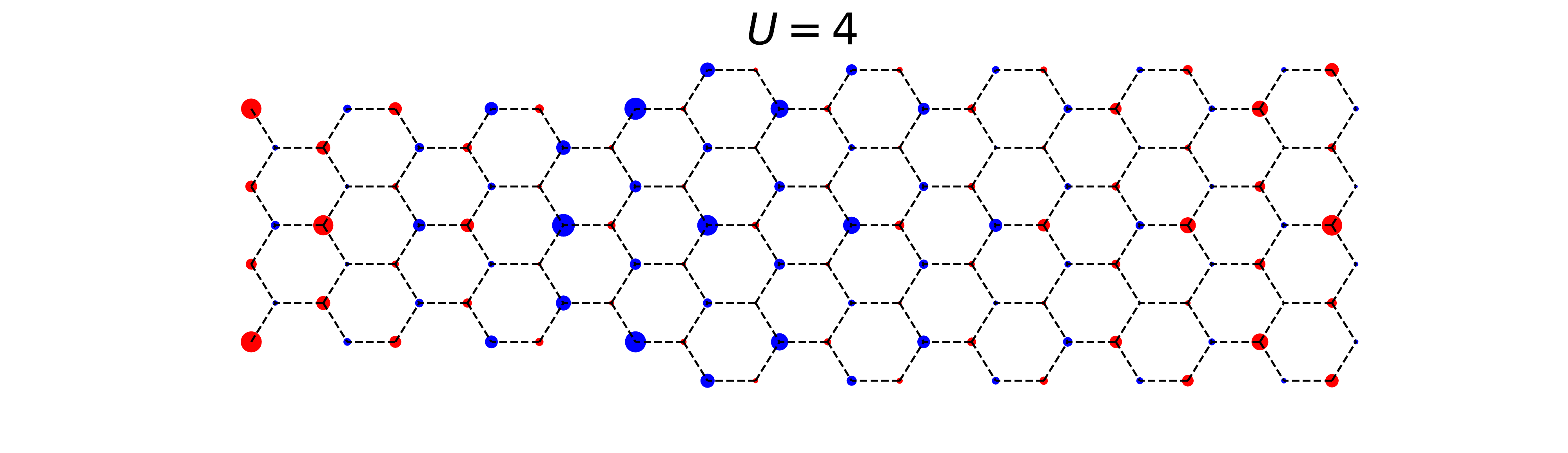}
    \caption{Density profile of lowest $k_x=0$ energy state for different values $U$, compared with the non-interacting case (i.e. $U=0$).}
    \label{fig:densities}
\end{figure}
\begin{figure}
    \centering
    \includegraphics[width=.8\textwidth]{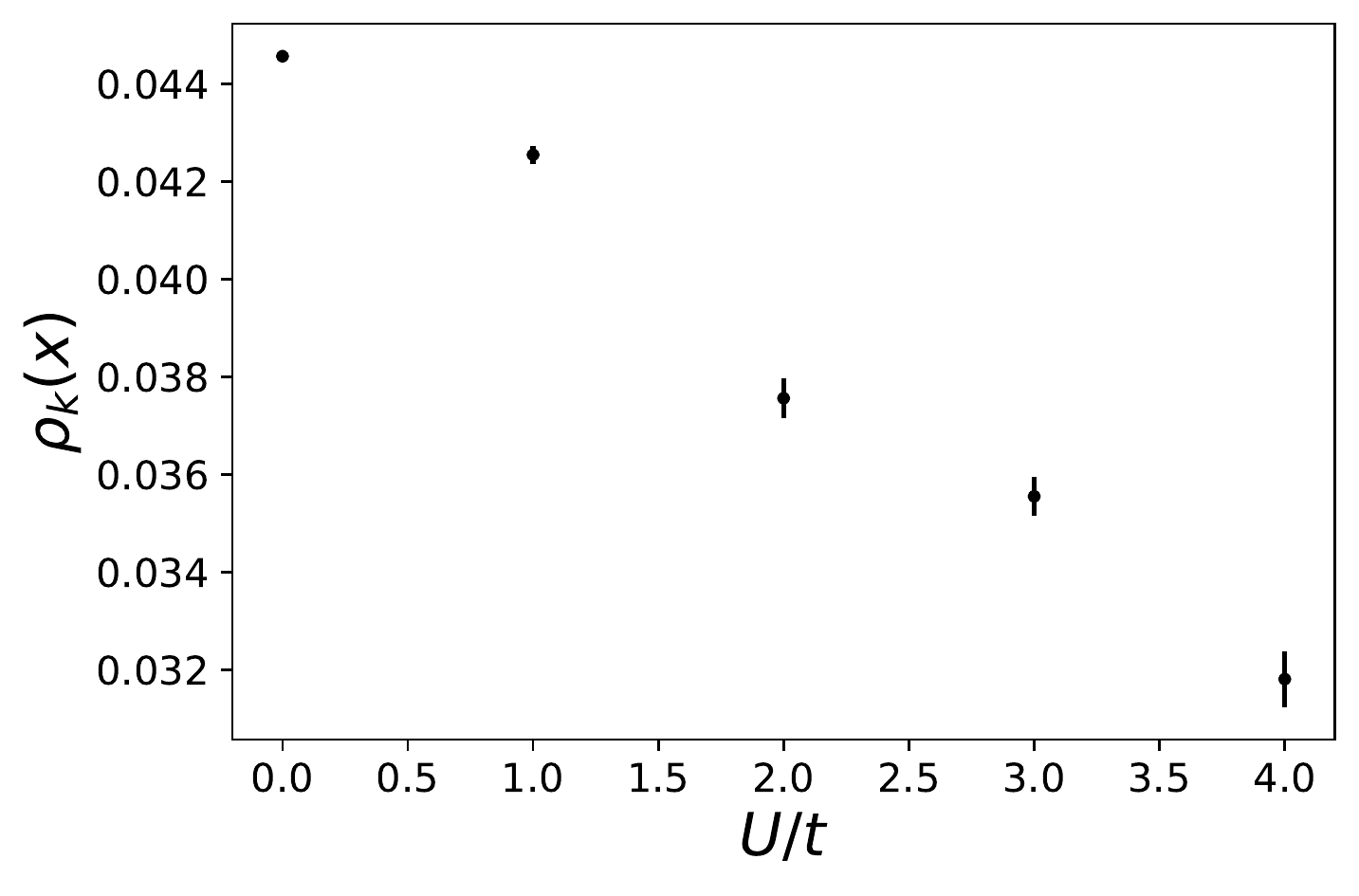}
    \caption{Wavefunction density $\rho_k(x)$ of the bottom- and left-most lattice site of our unit cell hybrid AGNR as a function of $U$.}
    \label{fig:amplitude vs U}
\end{figure}

We plot these densities for the localized state in Fig.~\ref{fig:densities} for different values of $U$.  We find that the changes in the densities vary only slightly as a function of $U$ and are practically indistinguishable in Fig.~\ref{fig:densities}.   In Fig.~\ref{fig:amplitude vs U} we concentrate on a specific lattice site, the bottom- and left-most site of the unit cell, and show how the density at this site varies as $U$ increases.  When $U=0$ this site is one of a four A sites that has a maximum probability for occupation compared to other A sites.  With increasing $U$ this density diminishes, but still remains the largest.  We find  a similar behavior with other high-density sites.  For sites with initially low probabilities at $U=0$, their densities slightly grow with increasing $U$.  However, the changes are too small to drastically change the general electron occupation profile. Thus the localization of this state persists as $U$ grows large, despite its growing energy.

Our results definitely show the strong dependence of the energy on $U$ within a finite volume.  The localization, however, is robust and persists in such environments.  A more definitive QMC investigation of this state would require repeated calculations of this system with more values of $\beta$ and number of timeslices $N_t$, as well as more unit cells, allowing for extrapolations to zero-temperature, to the continuum limit, and to the infinite volume (length), respectively.  We are actively pursuing this line of research.  

Still, the fact that these states remain localized for large values of $U$ within such an extreme finite volume bodes well for their potential utilization in advanced electronics, which by construction are finite in extent. 

\section{The symmetric-line limit\label{sect:symLine}}
We now consider the inclusion of a nearest-neighbor superconducting pairing term $\Delta$ to the Hamiltonian,
\begin{equation}\label{eqn:Hsym}
H_0=-\sum_{\langle i,j\rangle,\sigma}\left(t\  a^\dag_{i\sigma}a^{}_{j\sigma}+\Delta\  a^\dag_{i\sigma}a^\dag_{j\sigma}+\rm{h.c.}\right)+U\sum_x\left(n_{x\uparrow}-\frac{1}{2}\right)\left(n_{x\downarrow}-\frac{1}{2}\right)\ .
\end{equation}
The pairing term has the same symmetry properties as the hopping term, and in particular, the Hamiltonian remains invariant under time reversal.  Therefore the inclusion of this term does not change the topology of the system. 

As described in~\cite{Yang:2020lal,Ezawa2018,Miao:2019tng}, for example, when $\Delta$ has the same magnitude as the hopping parameter $t$, the onsite interaction term becomes quadratic in the number of creation and annihilation operators and therefore the spectrum of the system can be obtained by direct diagonalization.  We repeat the derivation for our system here.  We follow the conventions introduced in~\cite{Ezawa2018}.

Typically one uses a Bogoliubov-Valatin transformation \cite{Bogolyubov:1958km,Valatin:1958ja} in theories with pairing terms.  However, in this case, with an eye towards the interacting onsite term, we instead perform a  canonical transformation to a Majorana basis,
\begin{equation}\label{eqn:Majorana basis}
\begin{gathered}
a_{i\sigma}=\eta_{i \sigma}+i \gamma_{i \sigma}\ , \quad a_{i\sigma}^{\dagger}=\eta_{i \sigma}-i \gamma_{i \sigma}\ , \\
a_{j\sigma}=\gamma_{j \sigma}+i \eta_{j \sigma}\ , \quad a_{j\sigma}^{\dagger}=\gamma_{j \sigma}-i \eta_{j \sigma}\ ,
\end{gathered}
\end{equation}
where $i\in A$ sites and $j\in B$ sites.  The Hamiltonian in Eq.~\eqref{eqn:Hsym} then becomes
\begin{equation}
H=-2 i \sum_{\langle i, j\rangle \sigma}\left[\left (\Delta+t\right) \gamma_{i \sigma} \gamma_{j \sigma}+\left(\Delta-t\right) \eta_{i \sigma} \eta_{j \sigma}\right]-
U \sum_{x\in A\& B}\left(2 i \eta_{x \uparrow}\eta_{x \downarrow} \right)\left(2 i \gamma_{x \uparrow} \gamma_{x \downarrow}\right) .
\end{equation}
We now take the symmetric line limit by setting $\Delta=t$, thereby eliminating the $\eta$ Majorana fermions from the kinetic energy of the Hamiltonian above,
\begin{equation}\label{eqn:Hsym majorana}
H_{\rm sym}=-4 i t\sum_{\langle i, j\rangle \sigma} \gamma_{i \sigma} \gamma_{j \sigma}-
U \sum_{x\in A\&B}\left(2 i \eta_{x \uparrow}\eta_{x \downarrow} \right)\left(2 i \gamma_{x \uparrow} \gamma_{x \downarrow}\right) .
\end{equation}
Notice that the $\gamma$ Majorana fermions have a kinetic term similar to the original tight-binding Hamiltonian of Eq.~\eqref{eqn:H0}, but now with a hopping amplitude $4t$.  Indeed, when $U=0$ the dispersion for this system, when normalized by $4t$, is identical to the non-interacting dispersion shown in Fig.~\ref{fig:7_3.9_5 dispersion}.

Now consider the site-dependent operator $\hat d_j\equiv 2 i \eta_{j \uparrow}\eta_{j \downarrow}$.  One has that $[H_{\rm sym},\hat d_j]=0\ \forall\  j$.  Therefore, within Eq.~\eqref{eqn:Hsym majorana}, the term $2 i \eta_{x \uparrow}\eta_{x \downarrow}$ ($=\hat d_x$) can be replaced, in general, by a $c$-number $d_x$ (no hat symbol). However, Majorana operators $\eta$ have the property that $\eta^2=1/4$ which implies that $\hat d_x^2=1/4$.  Thus we can make the following replacement $\hat d_x\to d_x=\pm 1/2$ in Eq.~\eqref{eqn:Hsym majorana}.  This gives
\begin{equation}\label{eqn:Hsym 2}
H_{\rm sym}=-4 i t\sum_{\langle i, j\rangle \sigma}\gamma_{i \sigma} \gamma_{j \sigma}-
 2 i U\sum_{x\in A\&B}d_x\left( \gamma_{x \uparrow} \gamma_{x \downarrow}\right) .
\end{equation}
The equation above shows that in the symmetric line limit the $\eta$ Majorana fermions completely decouple from the theory.  They provide a zero-energy topological flat band to the to the dispersion, independent of $U$, but as argued in~\cite{Ezawa2018} these states do not correspond to localized states.

The Hamiltonian in Eq.~\eqref{eqn:Hsym 2} is quadratic in the Majorana operators and therefore can be directly diagonalized once the coefficients $d_i$ are fixed.  In principle, given $N$ lattice sites, there are $2^N$ different possible combinations of $d_i$, all satisfying the flat band condition for the $\eta$ Majorana fermions but providing a different spectrum for the $\gamma$ Majorana fermions.  We consider two uniform solutions in this work, the first being the ferromagnetic solution with $d_i=1/2\ \forall\ i$ and the other the antiferromagnetic case where $d_i=1/2$ for $i\in A$ sites and $d_i=-1/2$ for $i\in B$ sites.  Lastly we consider a  random configuration where  $d_i=\pm 1/2$ is chosen randomly at each site $i$.

\subsection{Ferromagnetic configuration}
\begin{figure}[h!]
    \centering
    \includegraphics[width=.5\textwidth]{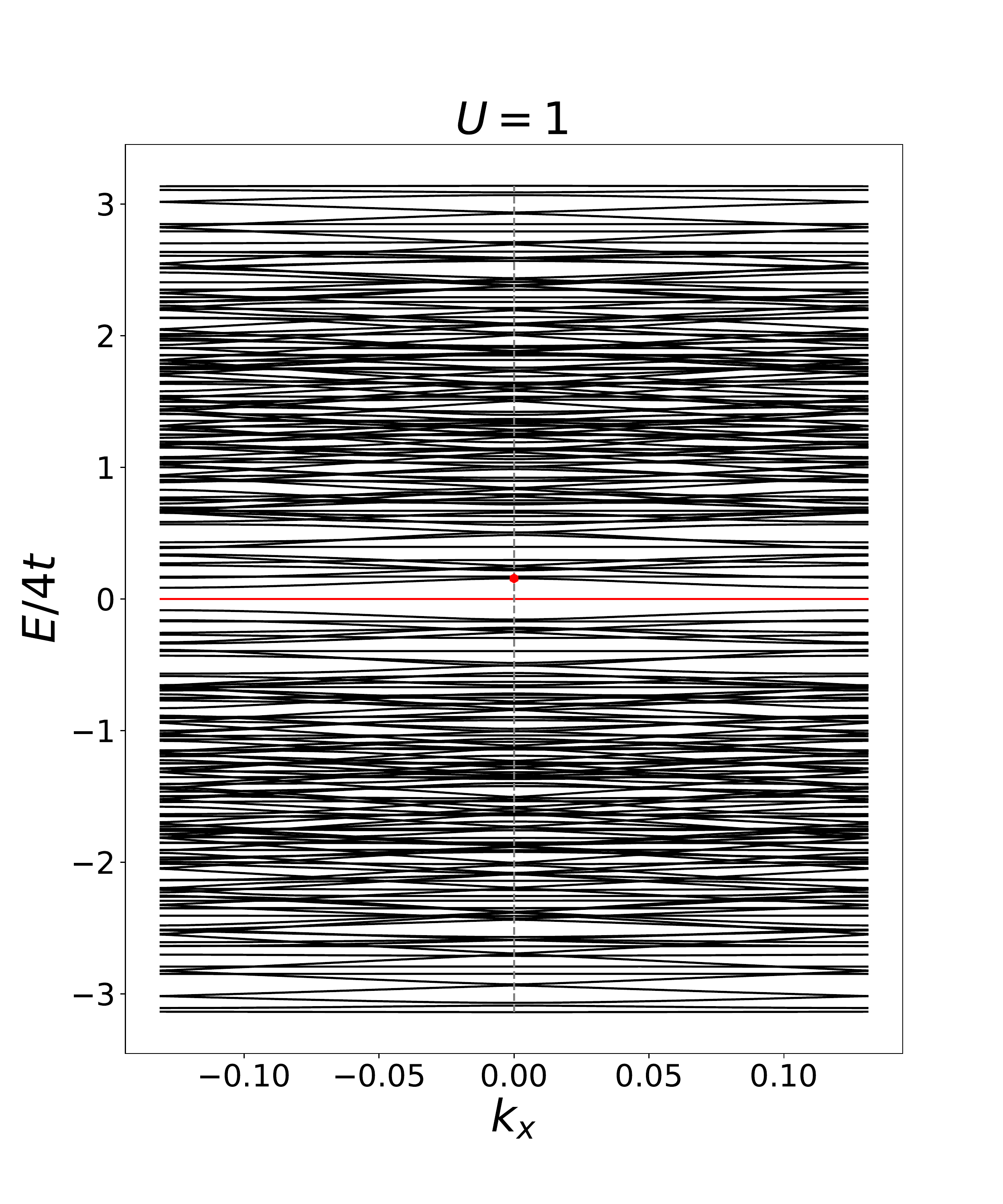}\includegraphics[width=.5\textwidth]{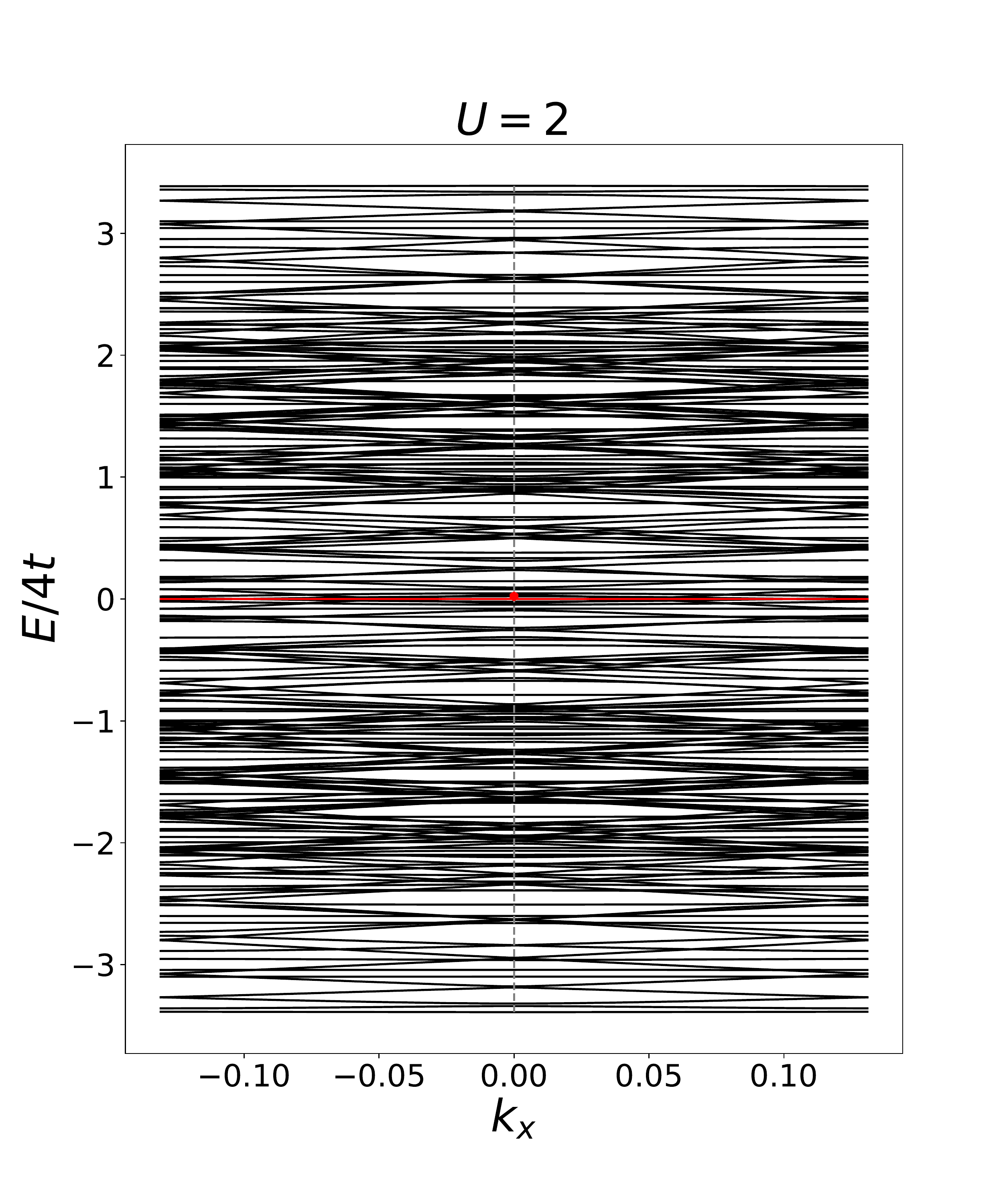}
    \includegraphics[width=.5\textwidth]{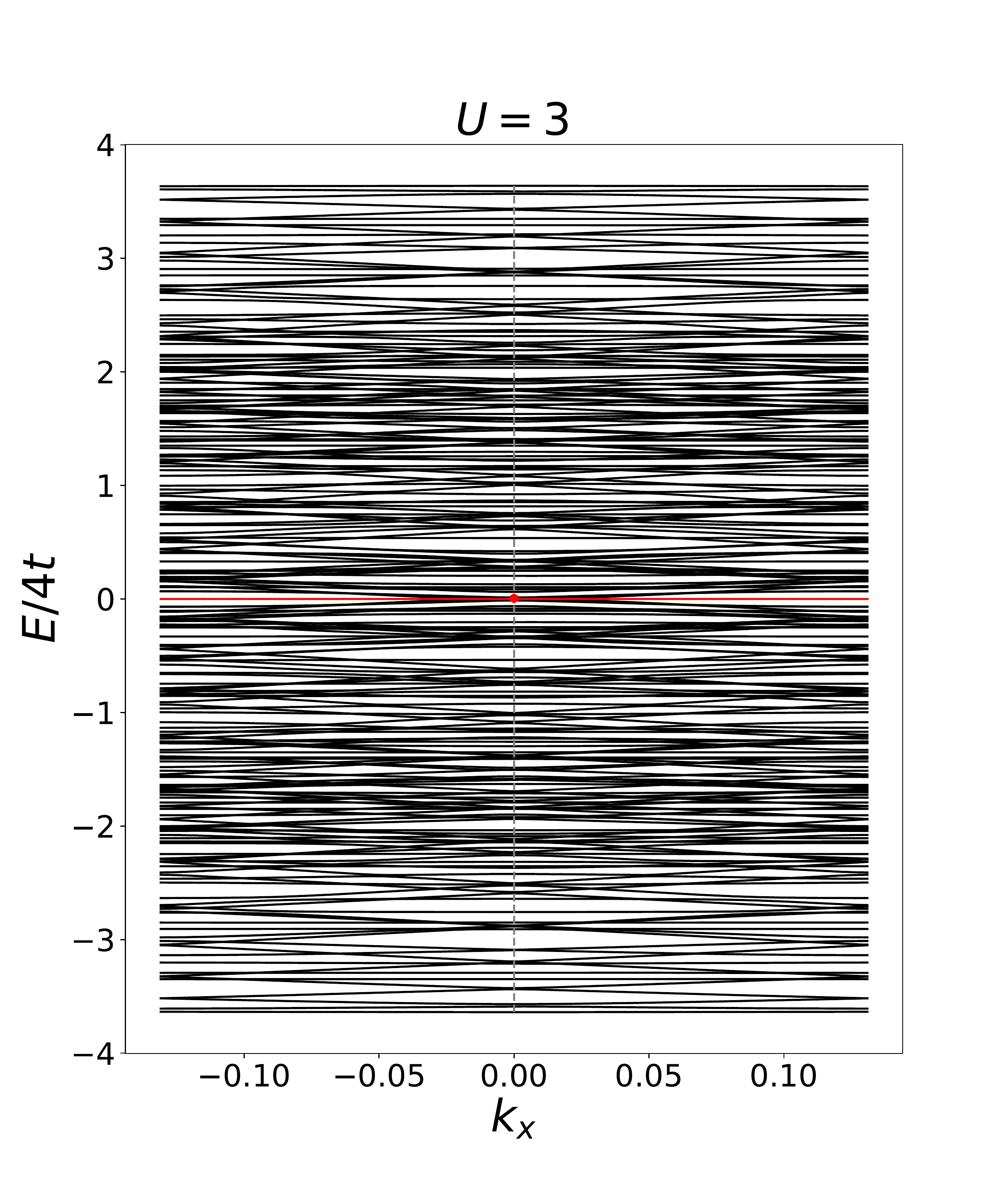}\includegraphics[width=.5\textwidth]{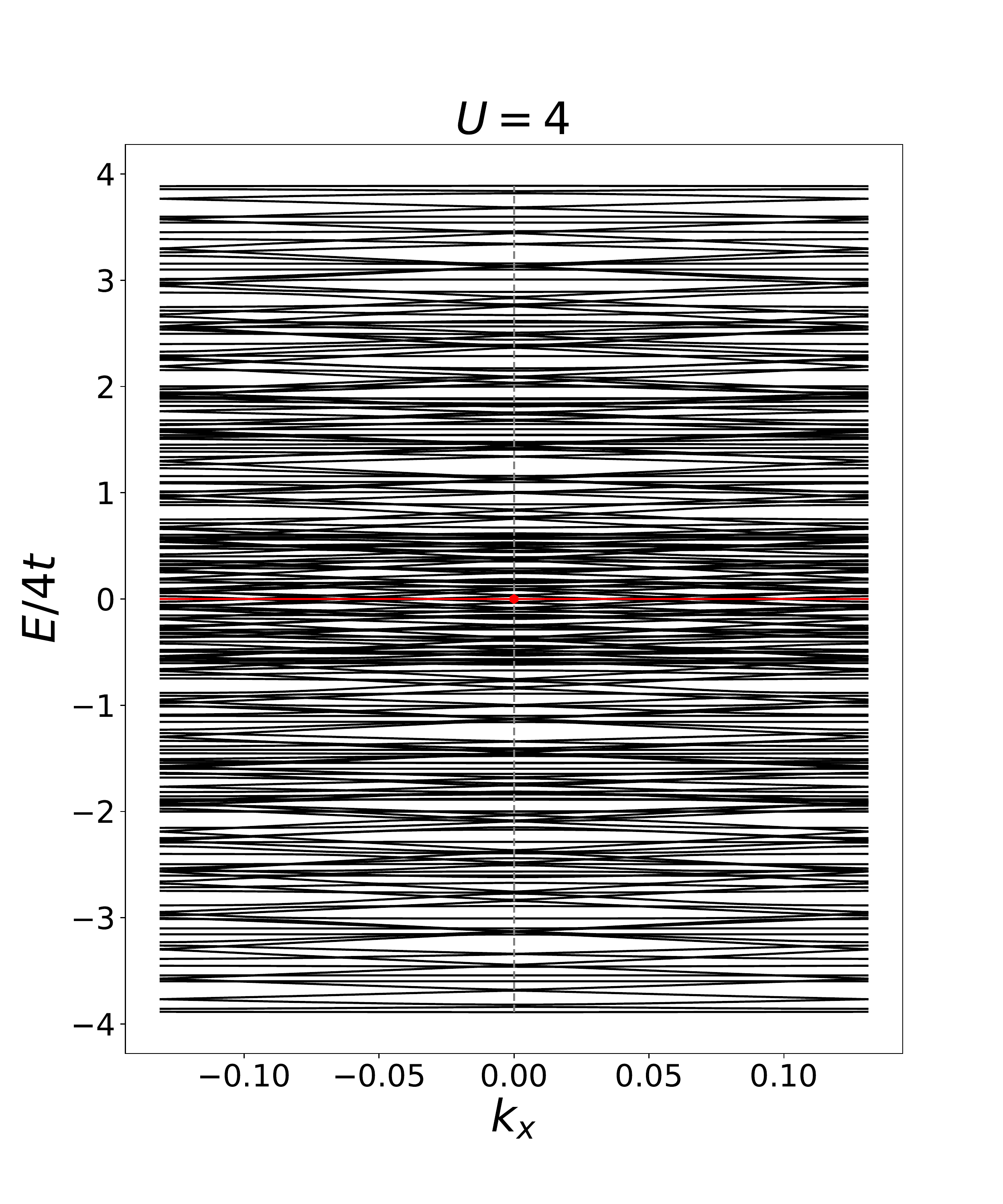}
    \caption{Ferromagnetic dispersion in the symmetric line limit for different values of $U$.  The red horizontal line is the flat band energy for the decoupled $\eta$ Majorana fermions.}
    \label{fig:symLine dispersion ferromagnetic}
\end{figure}

In this configuration we choose $d_i=1/2\ \forall\ i$.  Our results are identical if we instead chose $d_i=-1/2\ \forall\ i$.  We show the dispersion for this system for select values of $U$ in Fig.~\ref{fig:symLine dispersion ferromagnetic}.  In general the dispersion becomes quite dense and the separation between the lowest state and the next excited state diminishes as $U$ is increased.
\begin{figure}[t!]
    \centering
    \includegraphics[width=.8\textwidth]{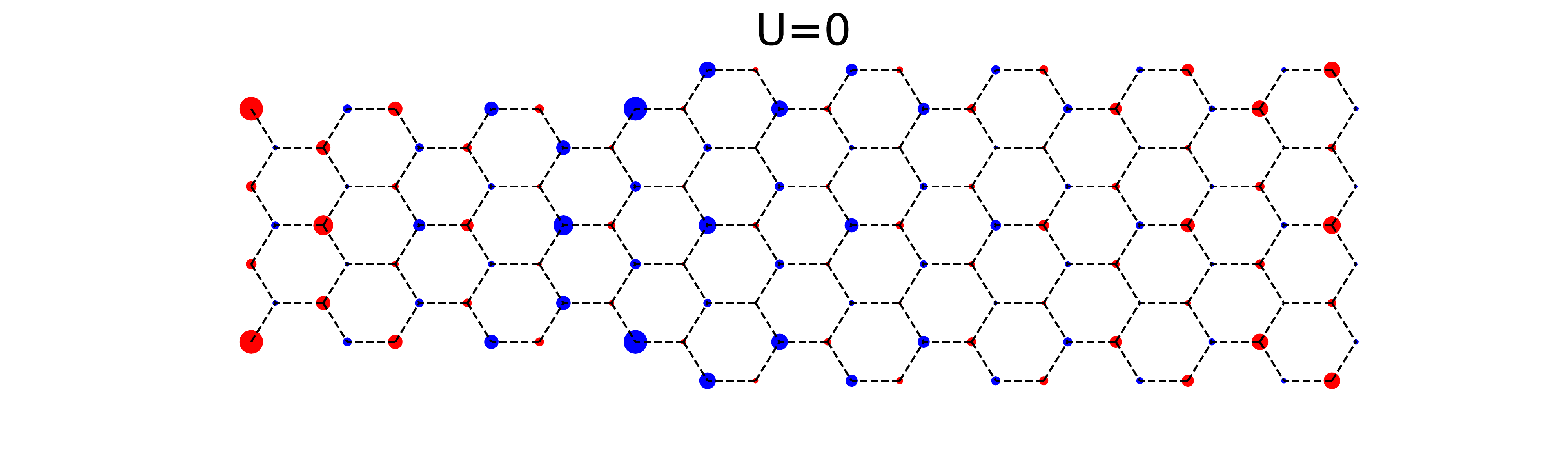}\\
    \includegraphics[width=.8\textwidth]{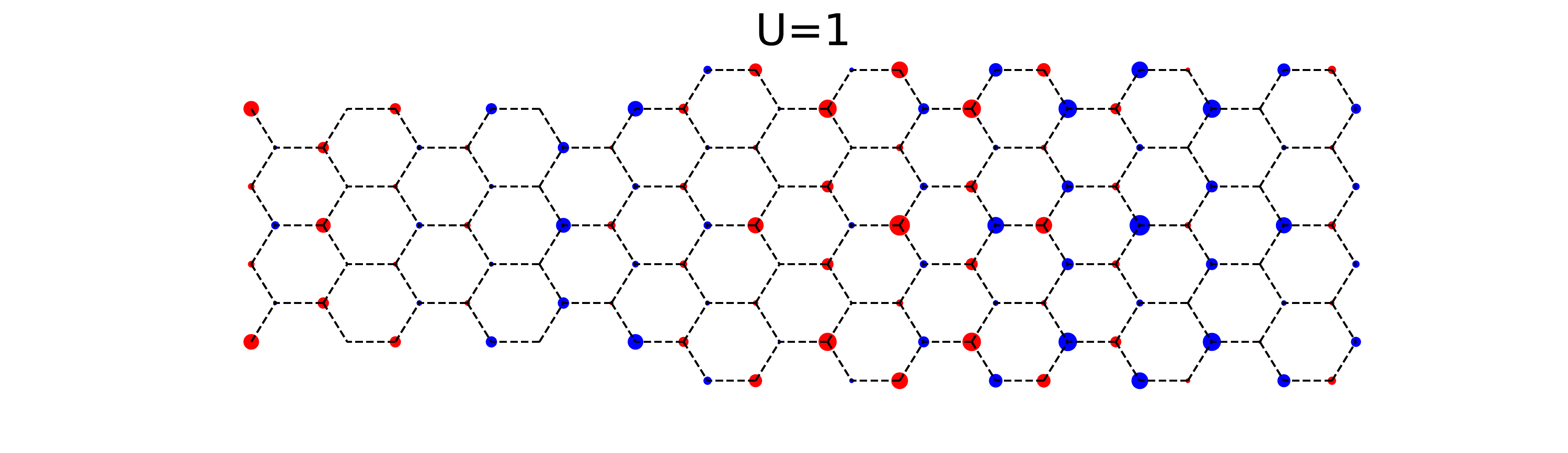}\\
    \includegraphics[width=.8\textwidth]{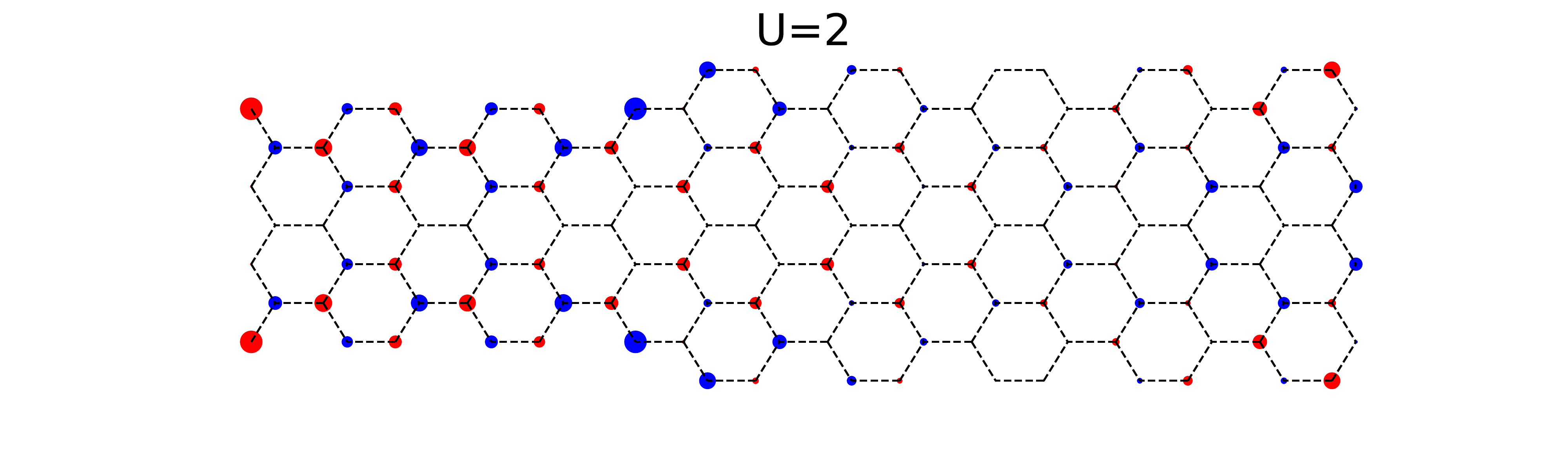}\\
    \includegraphics[width=.8\textwidth]{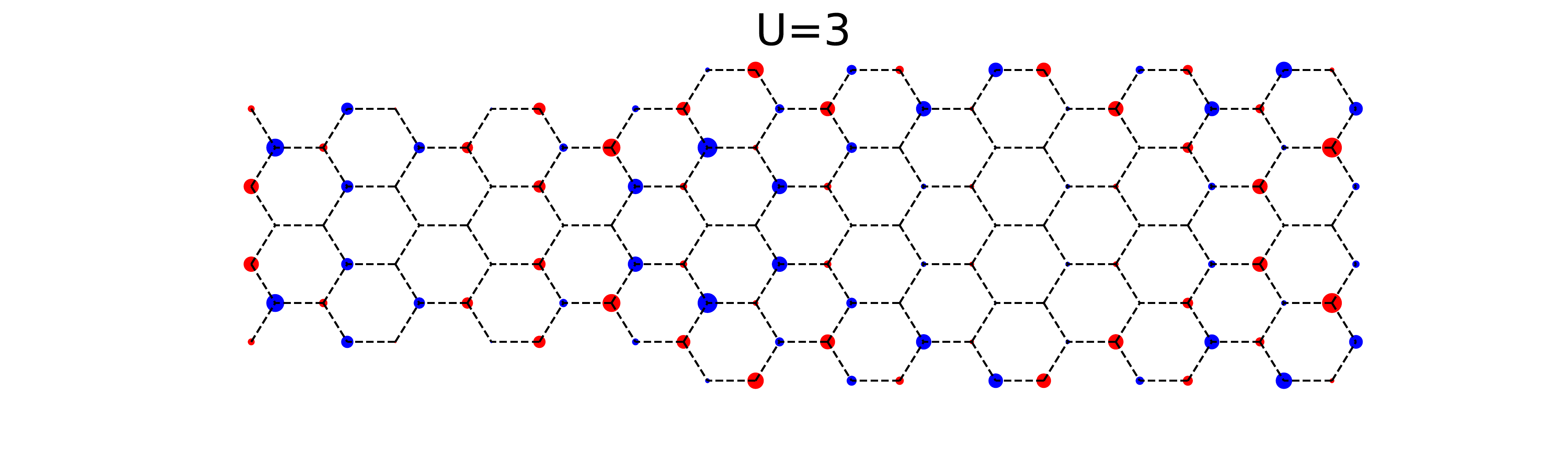}\\
    \includegraphics[width=.8\textwidth]{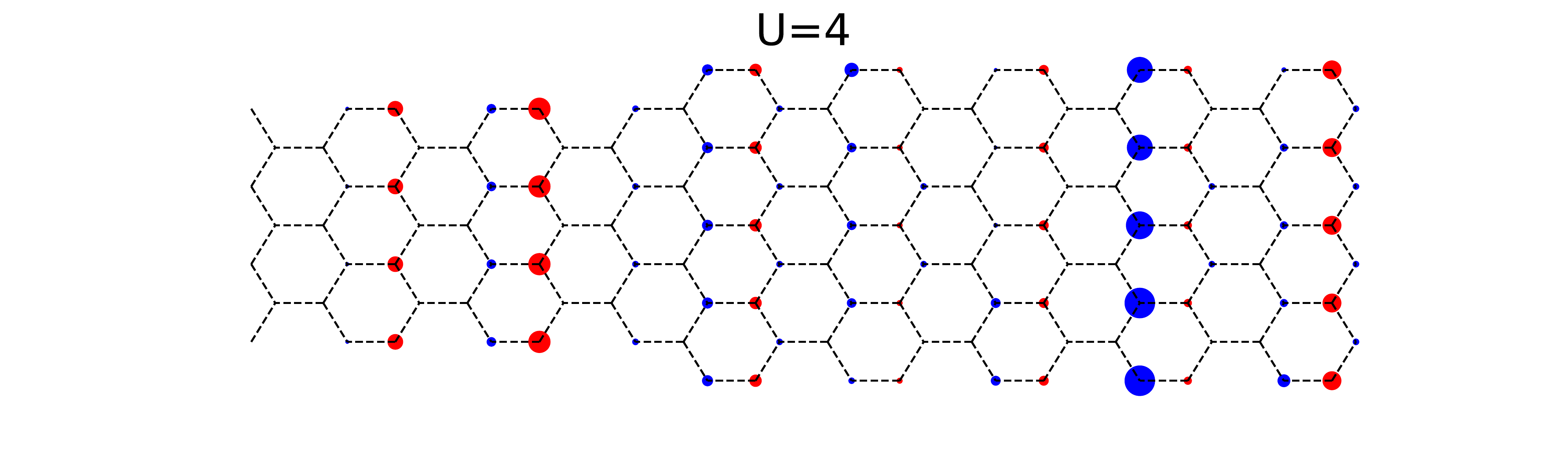}
    \caption{Density profile of the lowest $k_x=0$ energy state at the symmetric line limit for the ferromagnetic configuration for different values of $U$.  The result is the same for either spins $\sigma$.  Non-interacting case corresponds to $U=0$.}
    \label{fig:symLine densities ferromagnetic}
\end{figure}

The wavefunction densities for the lowest energy state are shown in Fig.~\ref{fig:symLine densities ferromagnetic}.  We find that this configuration exhibits no localization at $k_x=0$ for the large $U$s considered here, though we have confirmed that it is perturbatively recovered in the limit $U\to 0$.

\begin{figure}
    \centering
    \includegraphics[width=.8\textwidth]{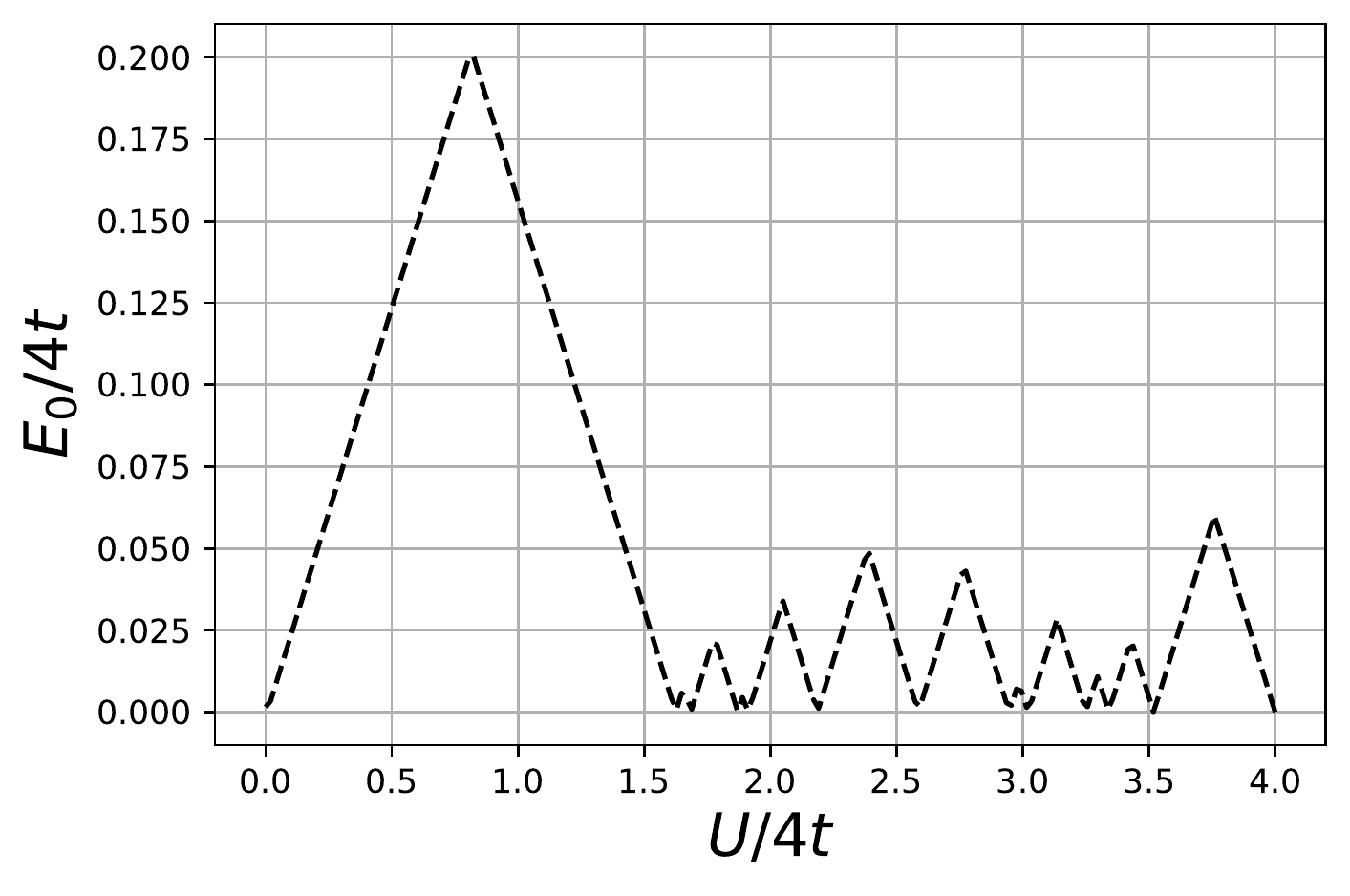}
    \caption{Dependence of the lowest positive energy $E_0$ as a function of $U$ in the ferromagnetic configuration.}
    \label{fig:U dependence ferromagnetic}
\end{figure}

Finally, the  energy $E_0$ of the lowest state has a complicated dependence on the interaction term $U$, as is shown in Fig.~\ref{fig:U dependence ferromagnetic}.

\subsection{Antiferromagnetic configuration}
\begin{figure}[h!]
    \centering
    \includegraphics[width=.5\textwidth]{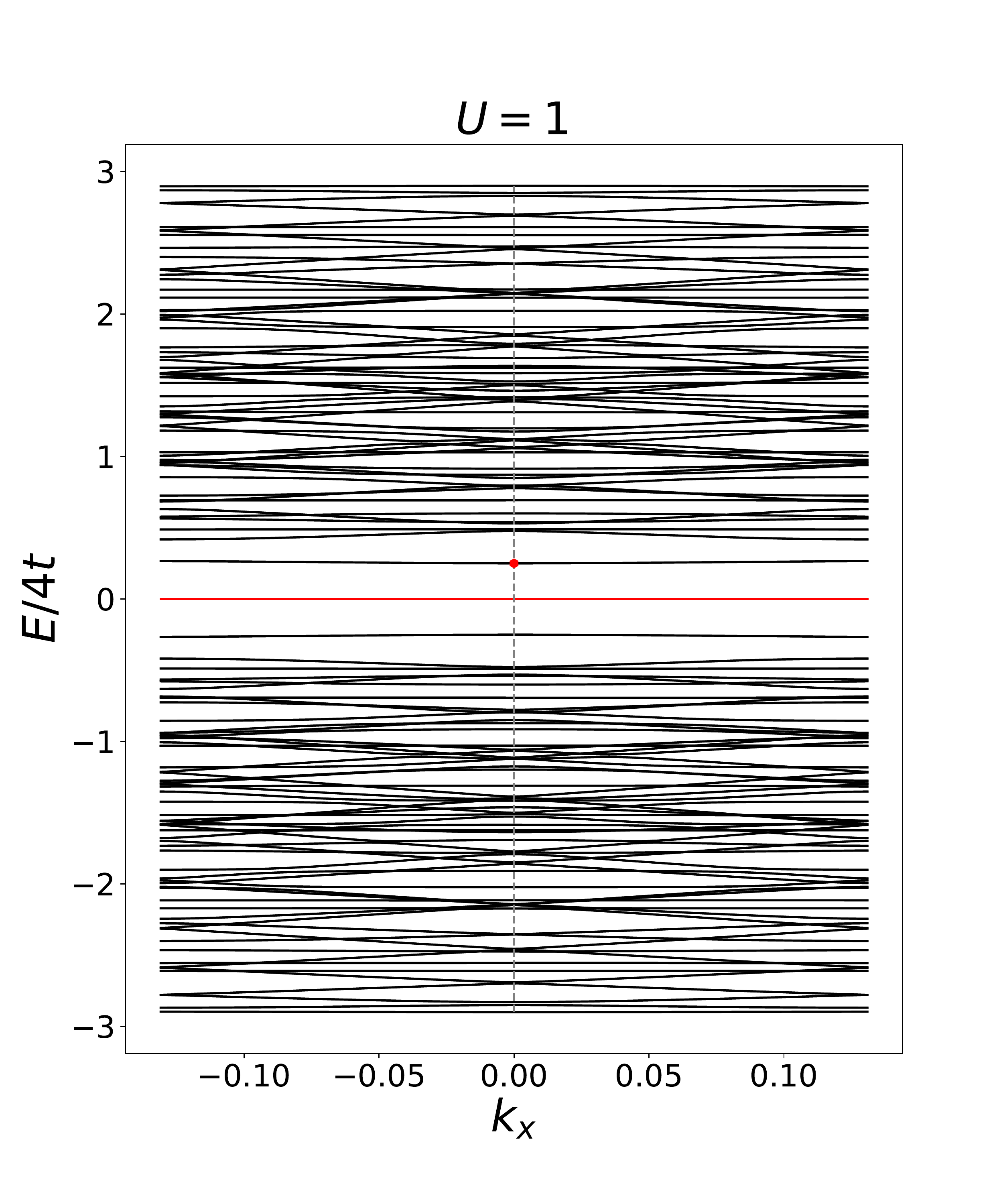}\includegraphics[width=.5\textwidth]{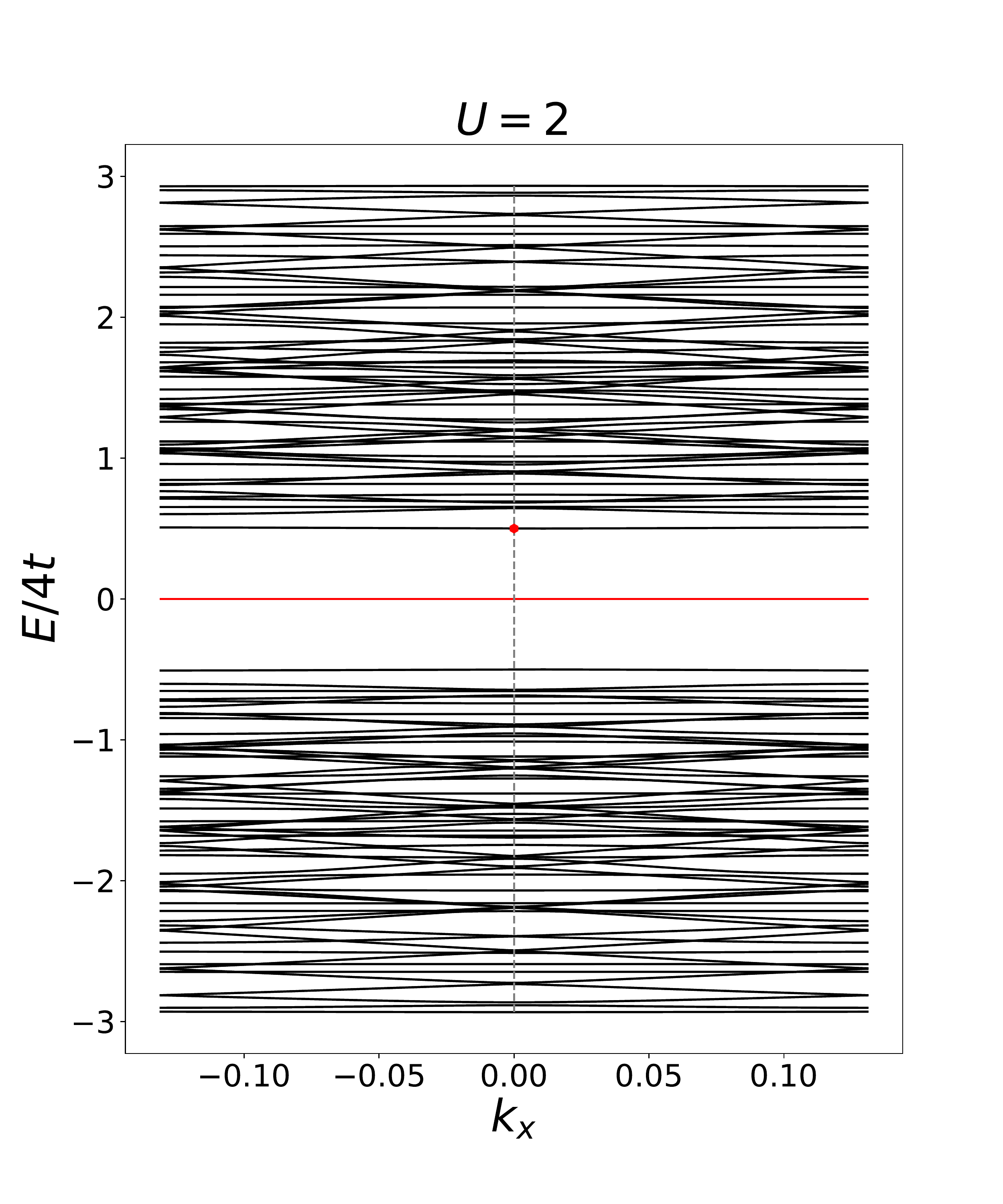}
    \includegraphics[width=.5\textwidth]{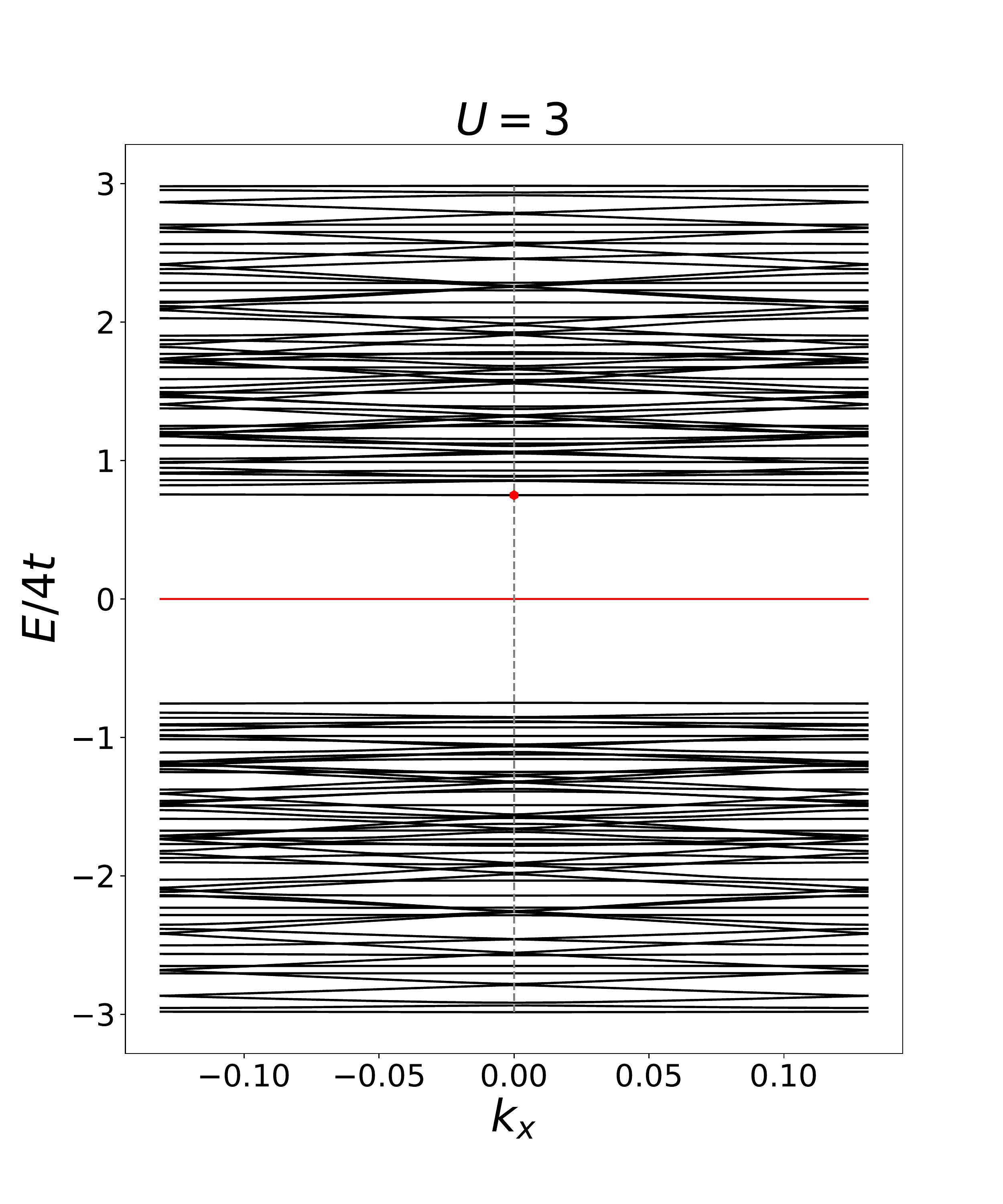}\includegraphics[width=.5\textwidth]{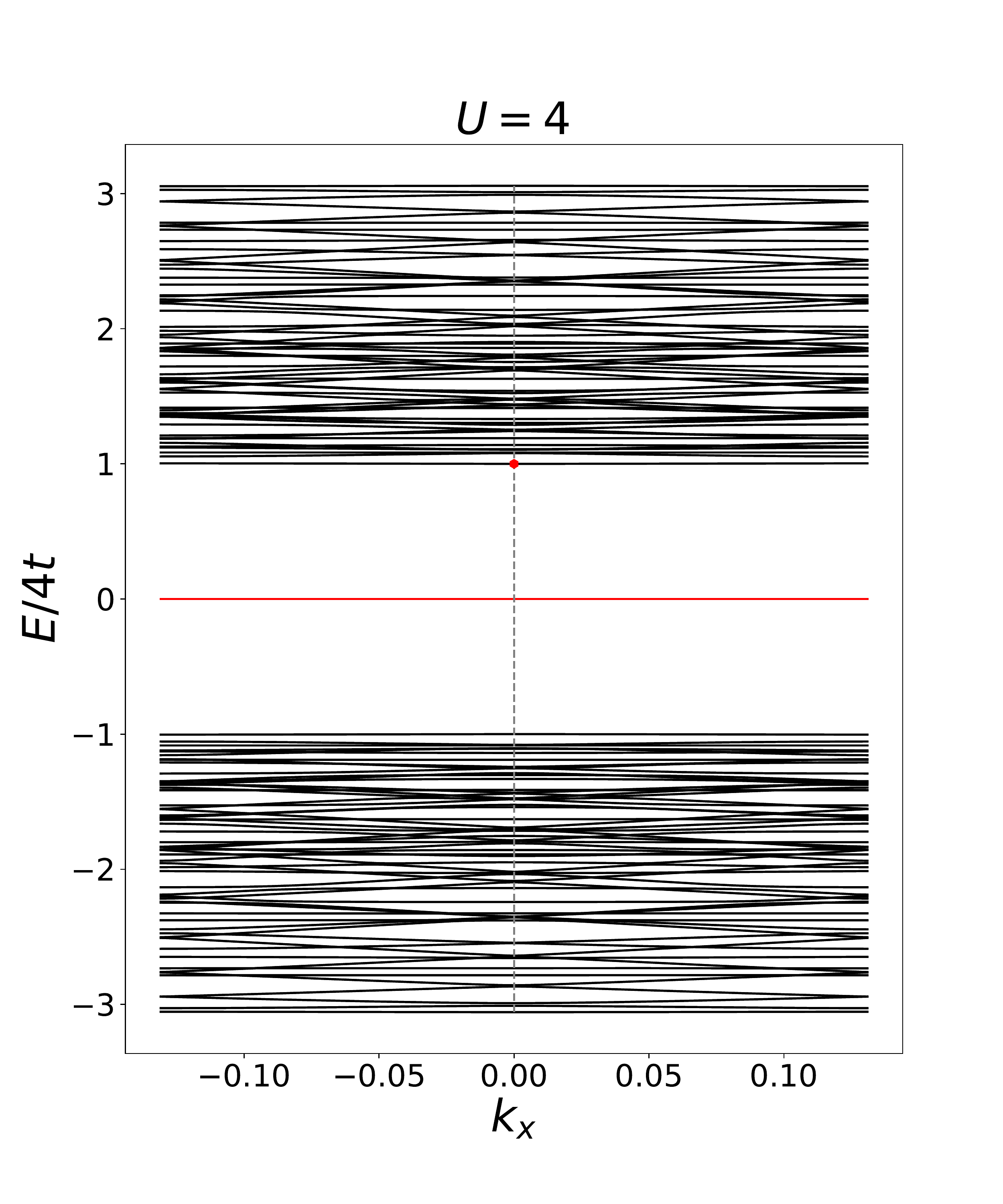}
    \caption{Antiferromagnetic dispersion in the symmetric line limit for different values of $U$.  The red horizontal line is the flat band energy for the decoupled $\eta$ Majorana fermions. }
    \label{fig:symLine dispersion}
\end{figure}

\begin{figure}
    \centering
    \includegraphics[width=.8\textwidth]{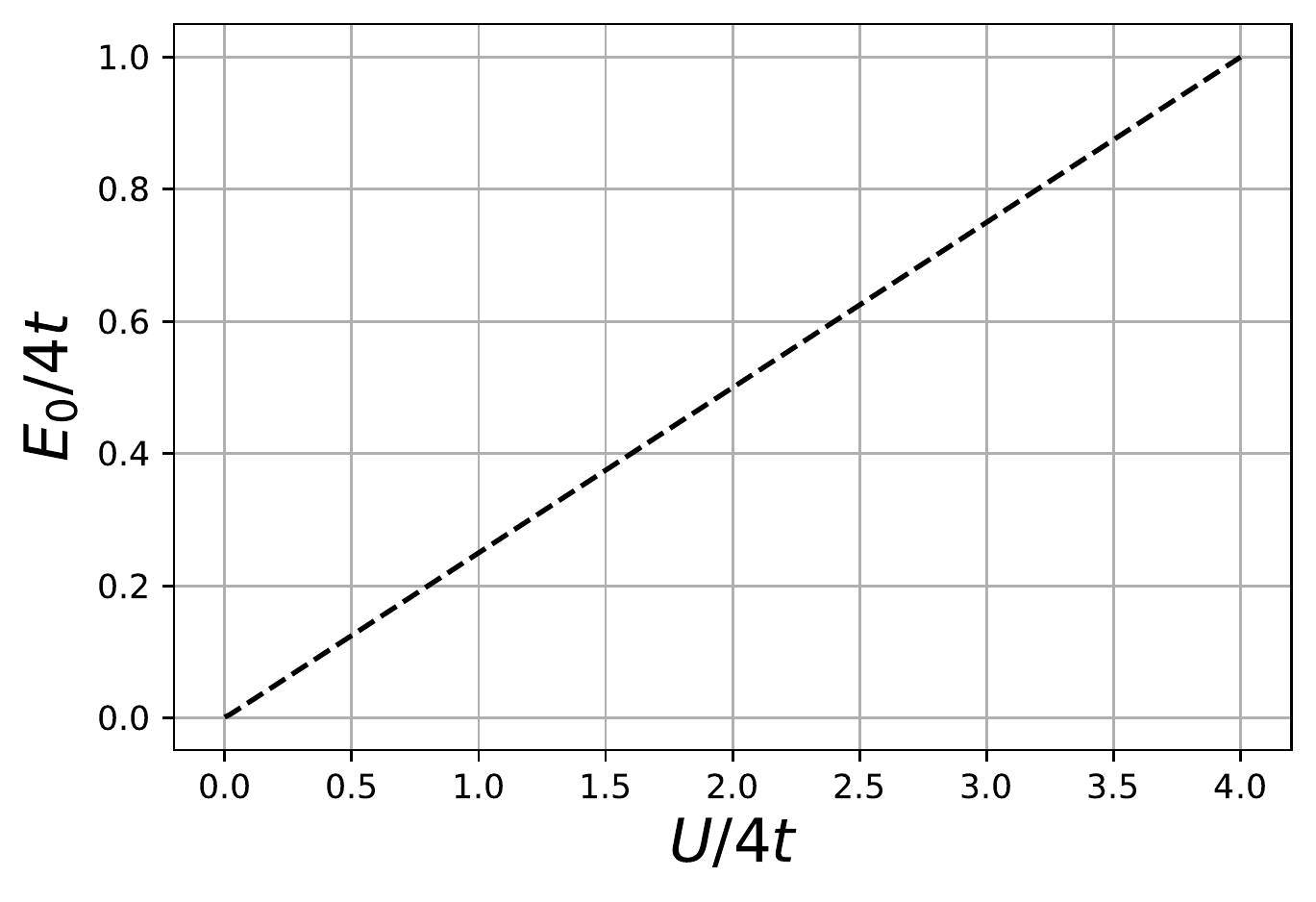}
    \caption{Dependence of the lowest positive energy $E_0$ as a function of $U$ in the antiferromagnetic configuration.}
    \label{fig:U dependence}
\end{figure}

Fig.~\ref{fig:symLine dispersion} shows the dispersion of the hybrid ribbon at the symmetric line limit for select values of $U>0$ in the antiferromagnetic configuration.  
Notice that the lowest positive energy increases with larger $U$ and forms essentially a flat band solution.  Numerically we find a linear dependence of this energy on $U$, as shown in Fig.~\ref{fig:U dependence}.

For all $U$s investigated, the wavefunction densities of this state does not change and remains exactly the same as that of the non-interacting state shown in Fig.~\ref{fig:79AGNR}.  Therefore this state remains localized, despite its energy having a linear dependence on $U$.  We conclude that the flat band that develops for $U>0$ is robust and is unaffected by interactions.

\subsection{Random configuration}
To a certain extent a random configuration of $d_i$s is similar to the antiferromagnetic configuration in that such a configuration has no long range order. Thus one might expect that the dispersion in the random configuration is similar to the antiferromagnetic case.  We find this to be true for values of $U$ as large as $U\lesssim 2$.  

\begin{figure}[h!]
    \centering
    \includegraphics[width=.5\textwidth]{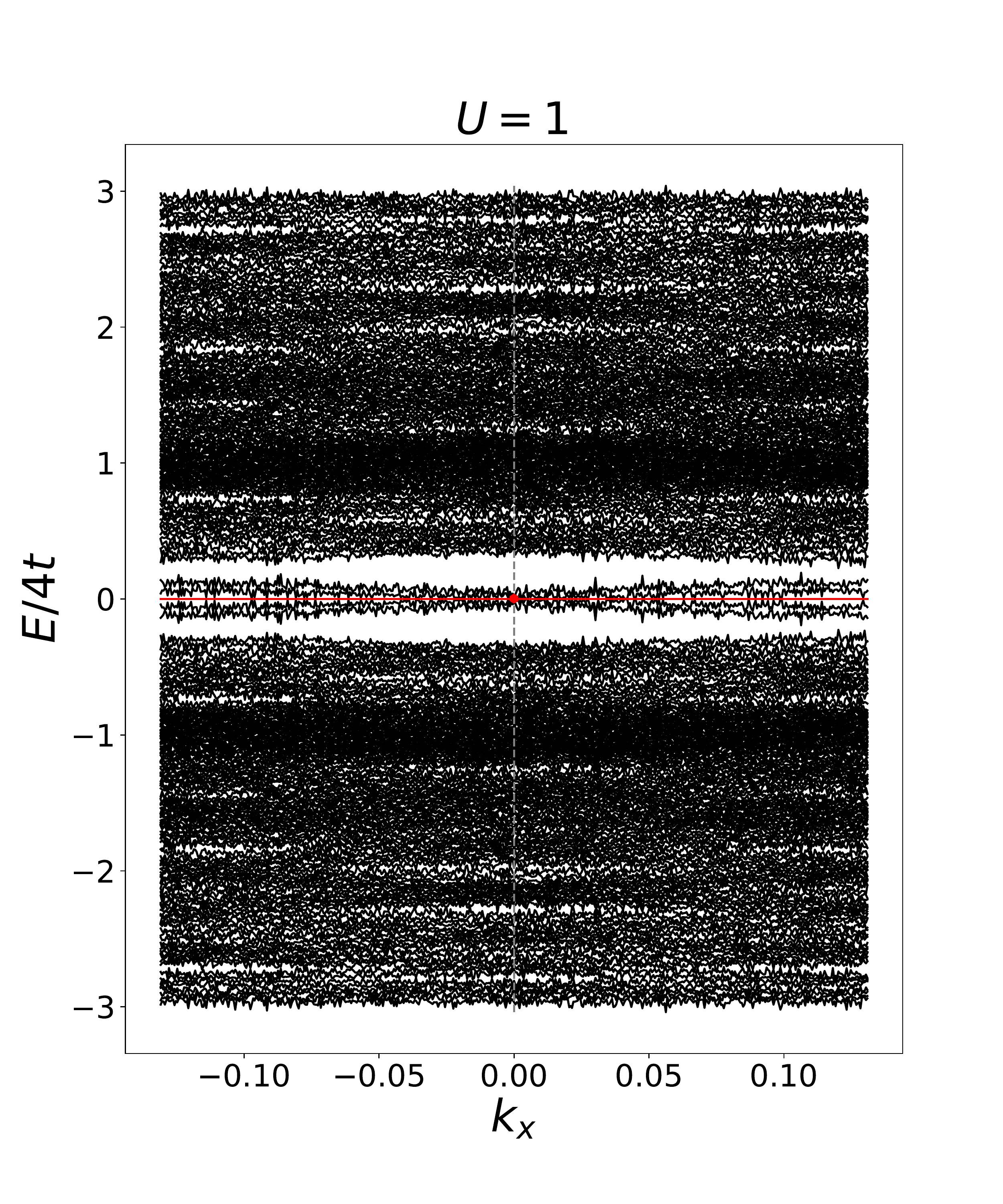}\includegraphics[width=.5\textwidth]{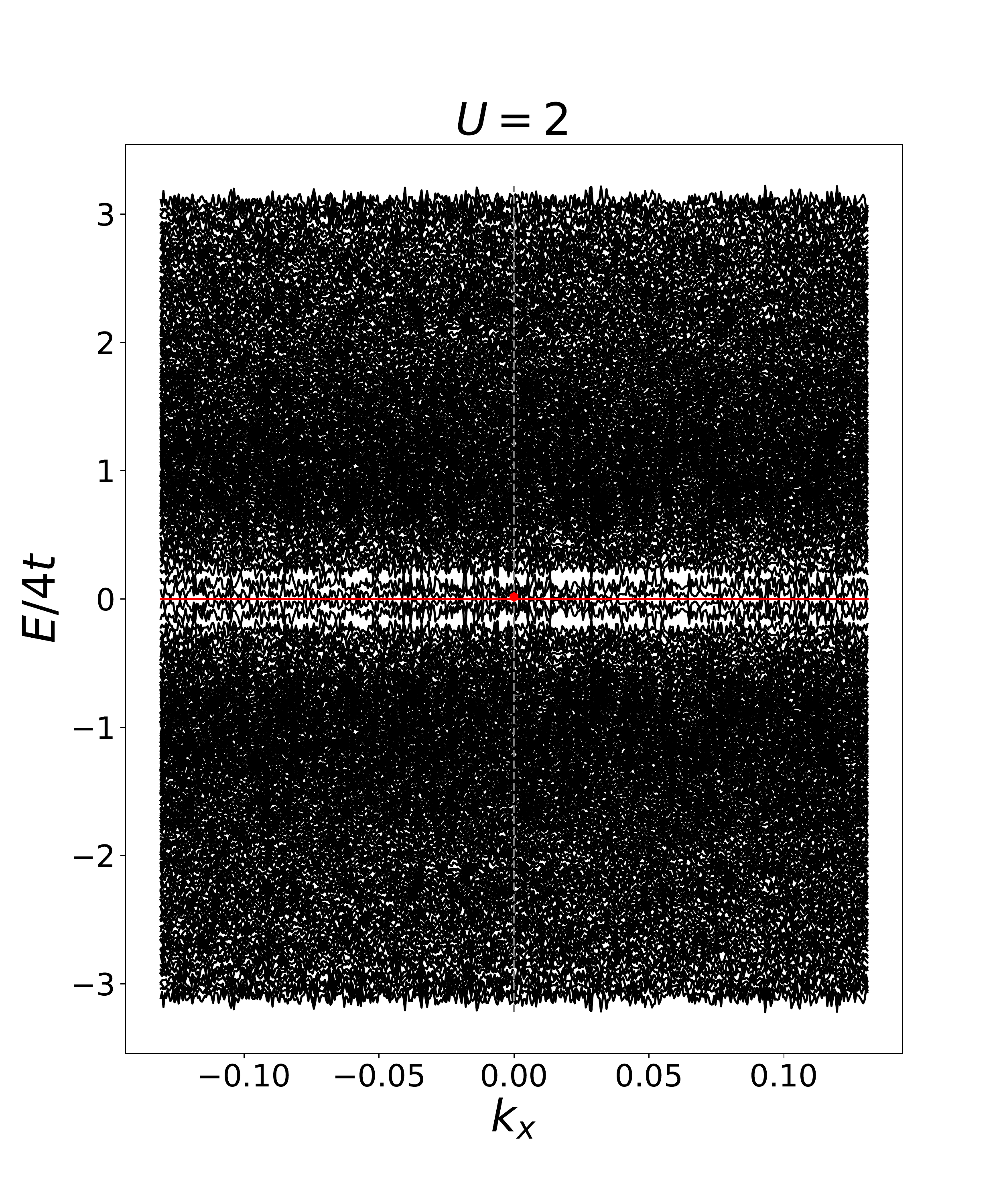}
    \includegraphics[width=.5\textwidth]{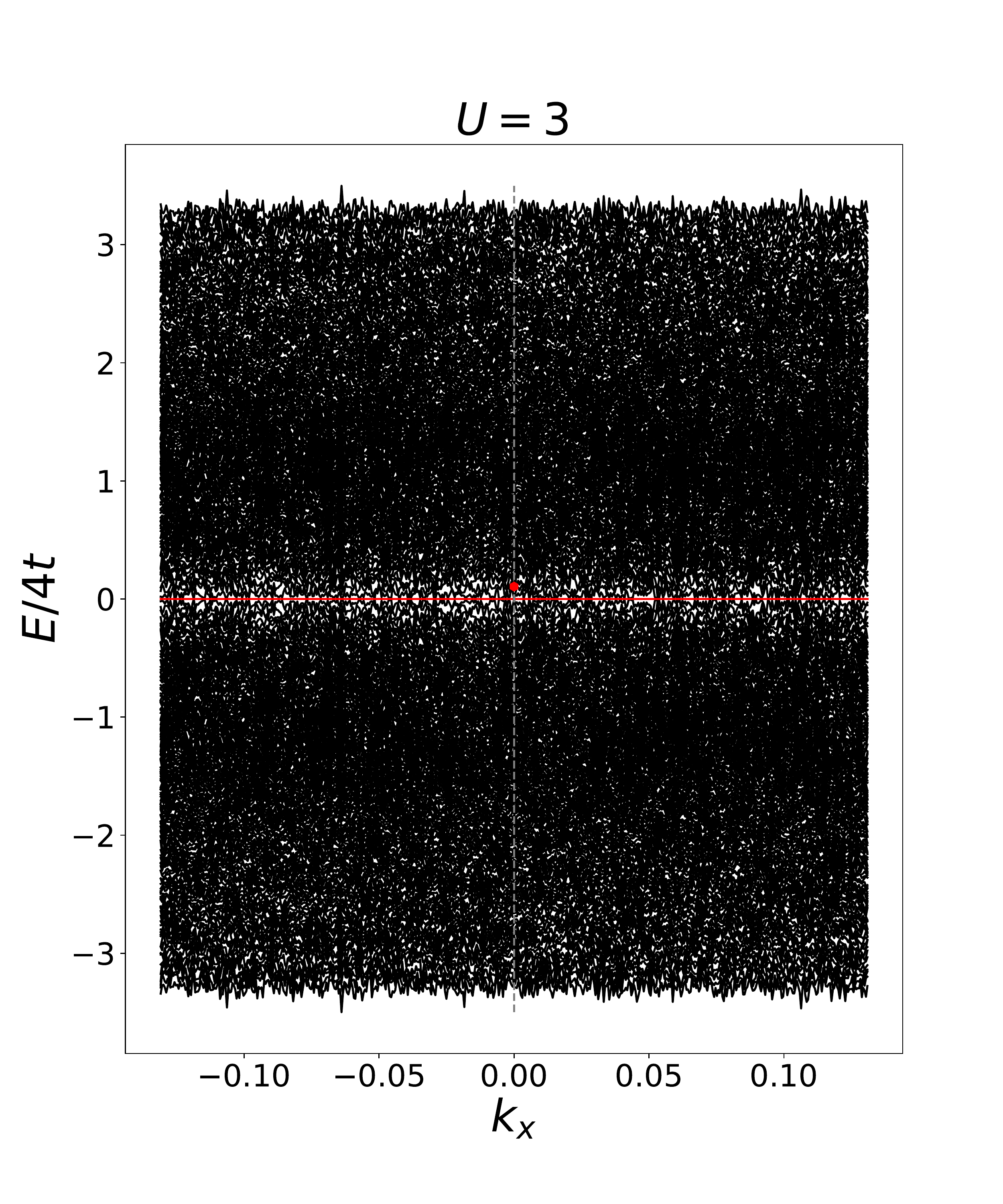}\includegraphics[width=.5\textwidth]{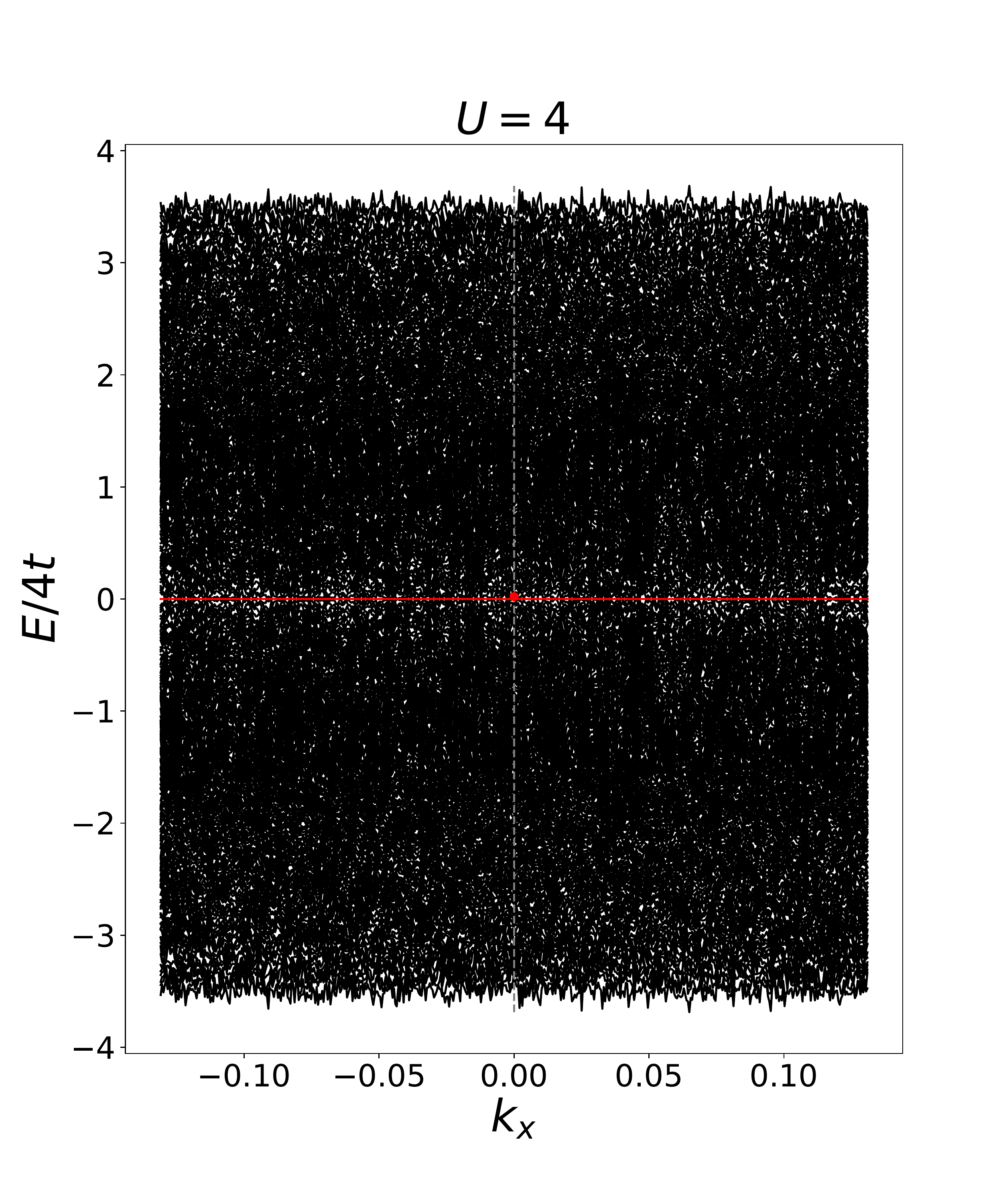}
    \caption{Dispersion at the symmetric line limit using random configurations.}
    \label{fig:symLine dispersion random}
\end{figure}

\begin{figure}
    \centering
    \includegraphics[width=.8\textwidth]{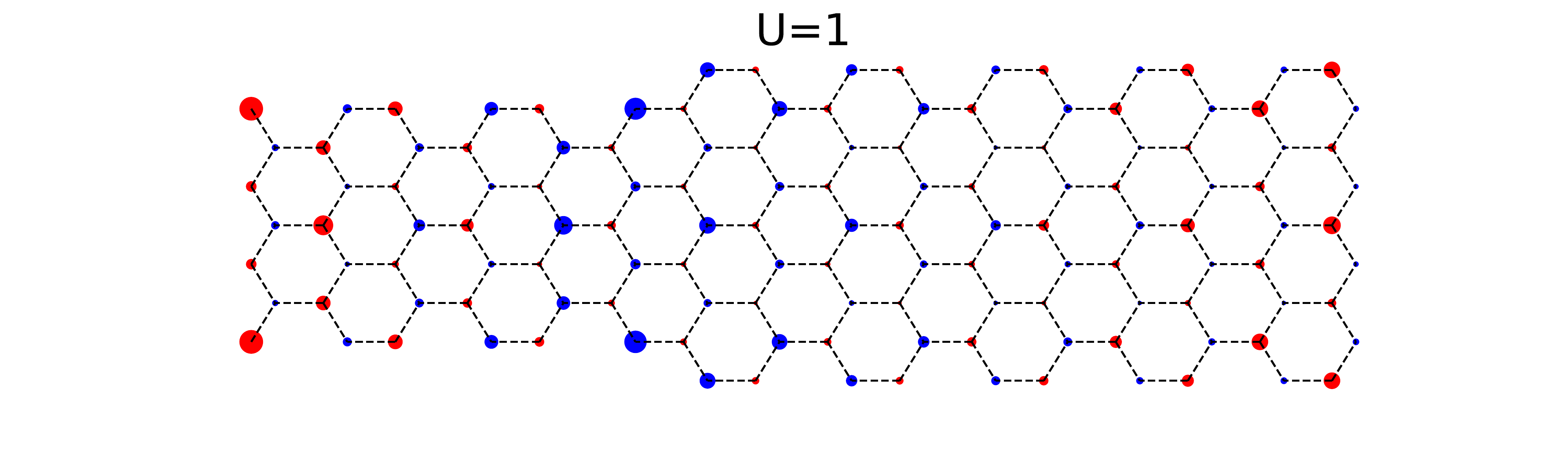}\\
    \includegraphics[width=.8\textwidth]{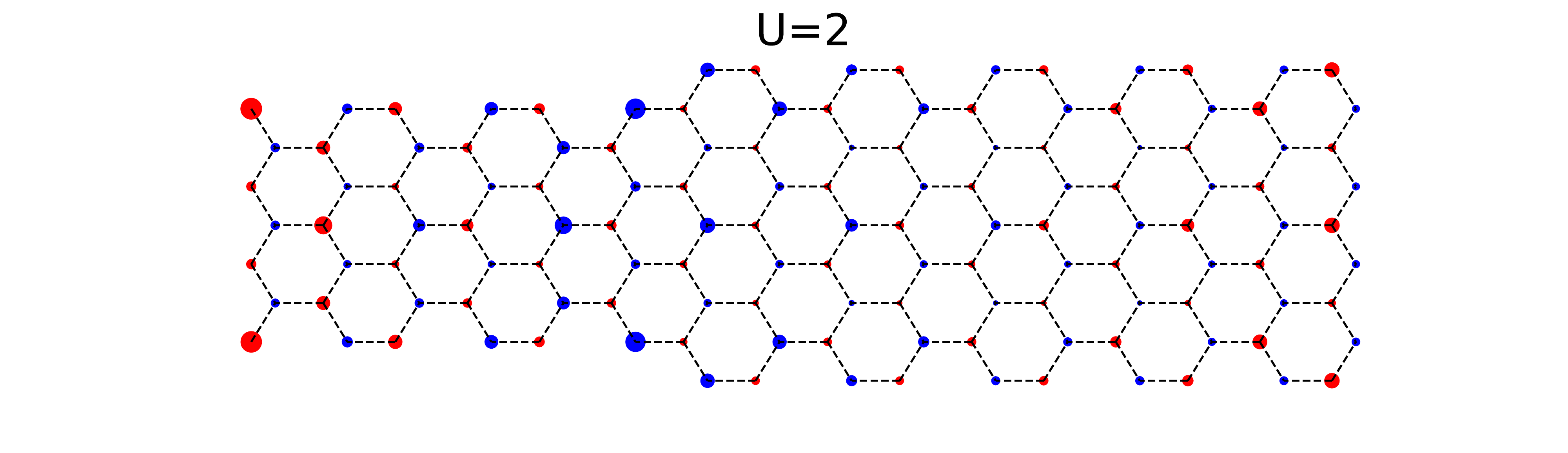}\\
    \includegraphics[width=.8\textwidth]{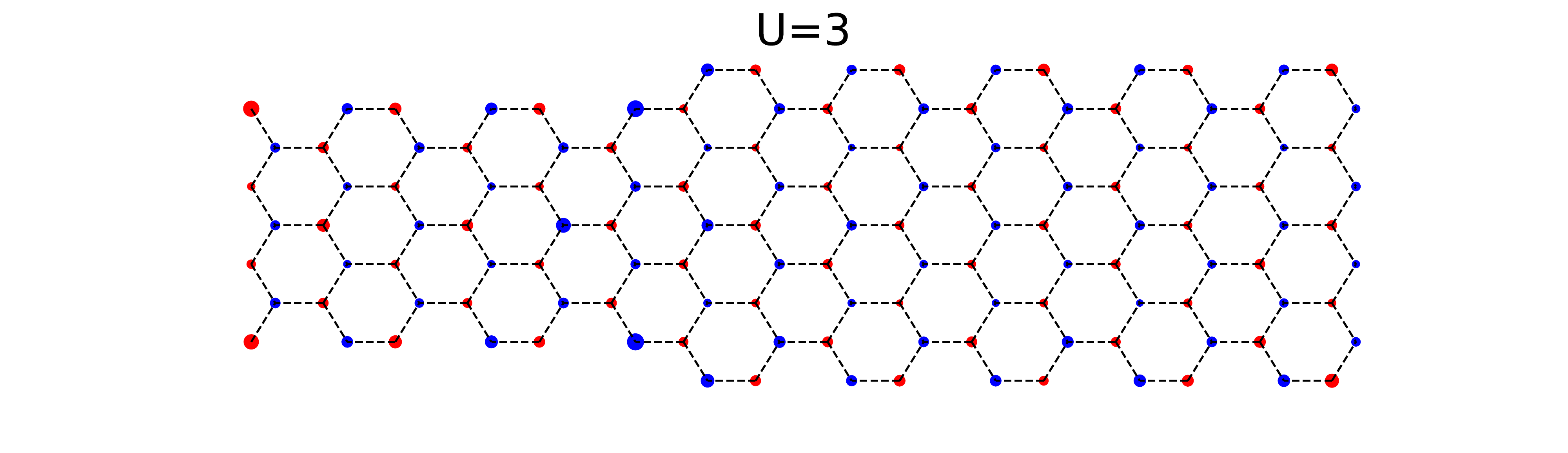}\\
    \includegraphics[width=.8\textwidth]{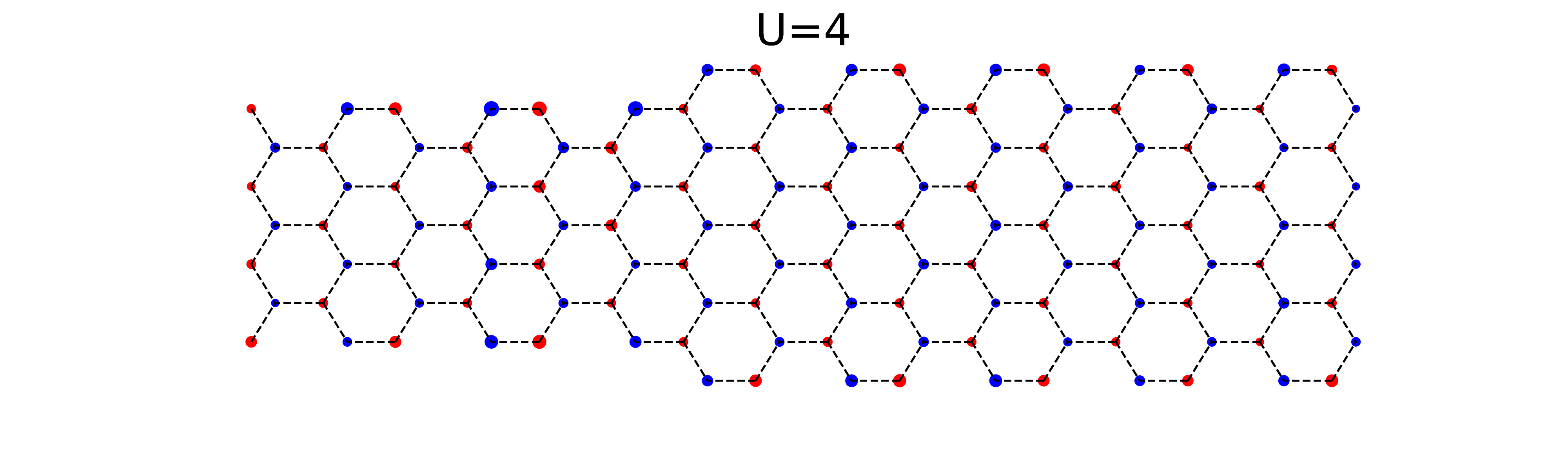}
    \caption{Density profile of the lowest $k_x=0$ energy state at the symmetric line limit using random configurations for different values of $U$. Localization can be seen in the cases with $U=1$ and $2$, but is lost for higher values.}
    \label{fig:symLine densities random}
\end{figure}

To see this, we first show in Fig.~\ref{fig:symLine dispersion random} the dispersion for different values $U$ using a single randomly sampled configuration in each case.  Not surprisingly, the dispersions becoming progressively dense and chaotic with increasing $U$.   To construct the accompanying wavefunction densities, we calculate 100 random configurations for each value of $U$ and average their wavefunction densities, the results of which are shown in Fig.~\ref{fig:symLine densities random}.  In this case the localization of the lowest state can be seen for $U=1$ and $U=2$.  However, for larger $U$ any analogies of the dispersion with the antiferromagnetic configuration is lost and localization is no longer present.

\section{Analogy with Domain-wall fermions\label{sect:dmf}}

Domain wall fermions were formulated originally by Kaplan~\cite{Kaplan:1992bt} as a way to circumvent the so-called Nielsen-Ninomiya no-go theorem~\cite{Nielsen:1981hk} in lattice gauge theory, which states that the number of left-handed chiral fermions must equal the number of right-handed chiral fermions in any discretized, local, Hermitian, and translationally invariant field theory.  Kaplan's formulation of domain wall fermions introduced an extra bulk dimension on top of the four spacetime dimensions, whereby a single fermion of one chirality was localized on the 4-d spacetime manifold (the domain wall where all the relevant physics occurs) of the 5-d space, and  another fermion of opposite chirality was constrained on the opposite 4-d domain wall.  In this manner, lattice gauge calculations utilizing domain wall fermions could simulate, in principle, an odd number of fermions with specific chirality by concentrating on one of the 4-d domain wall manifolds without violating the Nielsen-Ninomiya no-go theorem.  Chiral symmetry is still violated since the Ginsburg-Wilson equation remains non-zero in the bulk.  This manifests itself as a small overlap of the fermion wavefunctions in the bulk, and this in turn leads to a residual mass for each chiral fermion. As the bulk direction is extended, the overlap reduces leading to a vanishingly small residual mass.  Kaplan's formulation is actually valid for any theory in $2n+1$ dimensions, where $2n$ represents the spacetime dimension and the extra dimension represents the bulk.  

The localized states on opposite A/B sublattices, or chiralities, at the junctions of the topologically distinct ribbons offer a potential physical realization of these domain wall fermions in (quasi) $(0+1)+1$ dimensions.  Here the width of the junction is fixed and thus (quasi) zero dimensional, whereas the length of the ribbon between junctions represents the extra bulk dimension.  The additional dimension represents the temporal extent.  As the length between the separate ribbons is extended, the energy of the localized states approaches zero (in the non-interacting limit), which is analogous to the vanishing residual mass of the chiral fermions above as the bulk direction is extended.   A description of these chiral states is amenable to a corresponding effective field theory which we are currently developing.

\section{Conclusions\label{sect:conclusions}}
Localized states at the junction of topologically distinct nanoribbons offer promising avenues in constructing advanced electronics and potentially provide a means for topological, fault-tolerant quantum computing.  Central to this idea is the stability of such states not just to slight perturbations, but to large electron correlation effects.  In principle SPT provides this stability, but only in the limit of infinitely long ribbons where SPT invariance is manifest.  In a finite volume this protection is not guaranteed, and as such, the stability of such states comes into question when electron correlations become large.

In this paper we investigated the stability of the (nearly) zero-mode localized states in a finite 7/9 hybrid nanoribbon with periodic boundary conditions under the influence of temperature and electron-electron interactions.  We investigated two scenarios, one where we considered just the Hubbard model at half-filling and performed QMC simulations for a range of $U$ that included the strongly interacting regime.  We then introduced to the Hubbard model a nearest neighbor superconducting term whose parameter was tuned to the so called symmetric line limit. In this limit, when transforming to a Majorana basis, we could calculate the single-particle spectrum and wavefunctions exactly for any value of $U$.   Provided that we concentrate on the antiferromagnetic configuration in the latter case, we found that in both cases the energy of the localized states increased with larger $U$, but remained the lowest energy state regardless.  More importantly, we found that the localization of the states persisted at the junctions, indicating that this feature is robustly maintained in the strongly interacting, finite volume regime.  These findings enhance the possibility of using these systems for manufacturing novel electronic devices which are inherently finite in volume.

\begin{acknowledgments}
TL thanks Evan Berkowitz, Andrei Kryjevski, Johann Ostmeyer for enlightening discussions related to this work.  
This work was supported in part by the Chinese Academy of Sciences (CAS) President's International Fellowship Initiative (PIFI) 
(Grant No. 2018DM0034) and Volkswagen Stiftung (Grant No. 93562).
\end{acknowledgments}

\appendix

\section{Extracting site densities from QMC simulations\label{sect:densities}}
To extract the amplitudes for each site we first calculate site-dependent spatial correlators of the form
\begin{equation}\label{eqn:site correlator}
   C_k(x,t)\equiv \langle a_x^{}(t)a_k^\dag(0)\rangle=\frac{1}{Z} \operatorname{Tr}\ \left[a_x^{}(t)a^\dag_k(0)e^{-\beta H}\right]\ ,
\end{equation}
where $Z=\operatorname{Tr}\left[e^{-\beta H}\right]$ and the trace is taken over the entire Fock space of the system.  Here $x$ refers to a particular site on the lattice and $k=(k_x,\kappa)$ is the momentum variable that corresponds to the state that we are interested in.  The creation operator $a^\dag_k$ is
\begin{equation}\label{eqn:ak}
    a^\dag_k=\frac{1}{N_u}\sum_{x_u,i}e^{-ik_x x_u}C_i^\kappa a^\dag_{x_u,i}\ ,
\end{equation}
where the sum is over $N_u$ locations of the unit cells located at positions $x_u$ and the ions $i$ within each unit cell.  The coefficients $C_i^{\kappa}$ are the non-interacting eigenvector components obtained from the diagonalization of the tight-binding Hamiltonian.  For the low-energy localized state, we have that $k_x=0$ and choose $\kappa$ to correspond to the (non-interacting) eigenvector corresponding to this localized state.

By expressing the time-dependence in the right-hand side of Eq.~\eqref{eqn:site correlator} in the Heisenberg picture, 
\begin{displaymath}
a_x(t)=e^{-Ht}a_xe^{Ht}\ ,
\end{displaymath}
we can perform a spectral decomposition and determine the leading dependence of this correlator in the large time limit.  We find
\begin{equation}\label{eqn:scaling behavior}
    \lim_{1\ll t <\beta}\ C_k(x,t)= \langle \Omega|a_x|\Omega+k\rangle\langle \Omega+k|a^\dag_k|\Omega\rangle e^{-(\varepsilon_{\Omega+k}-\varepsilon_\Omega)t}+\ldots\ ,
\end{equation}
where the ellipsis represents terms that are exponentially suppressed.  The state $|\Omega\rangle$ and its associated energy $\varepsilon_\Omega$ represents the half-filling global ground state and global interacting energy minimum, respectively, and the state $|\Omega+k\rangle$ and associated energy $\varepsilon_{\Omega+k}$ is the state with an additional fermion with momentum $k$ above half filling and its corresponding interacting energy, respectively. The energy \emph{difference} $\varepsilon_{\Omega+k}-\varepsilon_\Omega\equiv E_k$ is exactly the interacting energy that we refer to in the manuscript. 
\begin{figure}
    \centering
    \includegraphics[width=.8\textwidth]{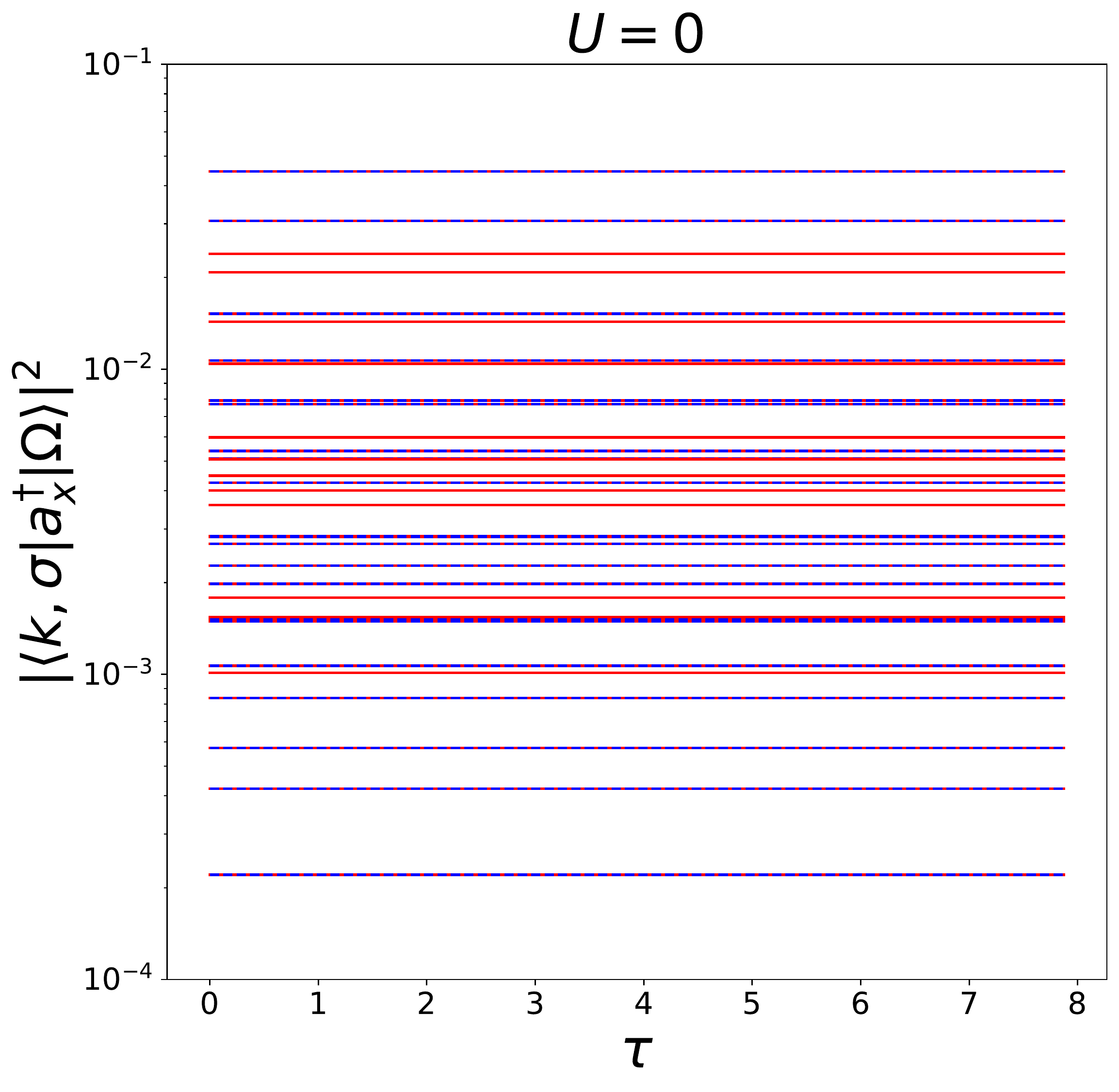}
    \caption{Non-interacting wavefunction densities $\rho_k(x,t)$ (labeled as $|\langle k,\sigma|a^\dag_x|\Omega\rangle|^2$ above) of the lowest energy localized state, as defined by Eq.~\eqref{eqn:density definition}.  The different lines correspond to different lattice sites $x$ and the red/blue coloring refer to A/B sites.}
    \label{fig:non interacting amplitudes}
\end{figure}

\begin{figure}
    \centering
    \includegraphics[width=.5\textwidth]{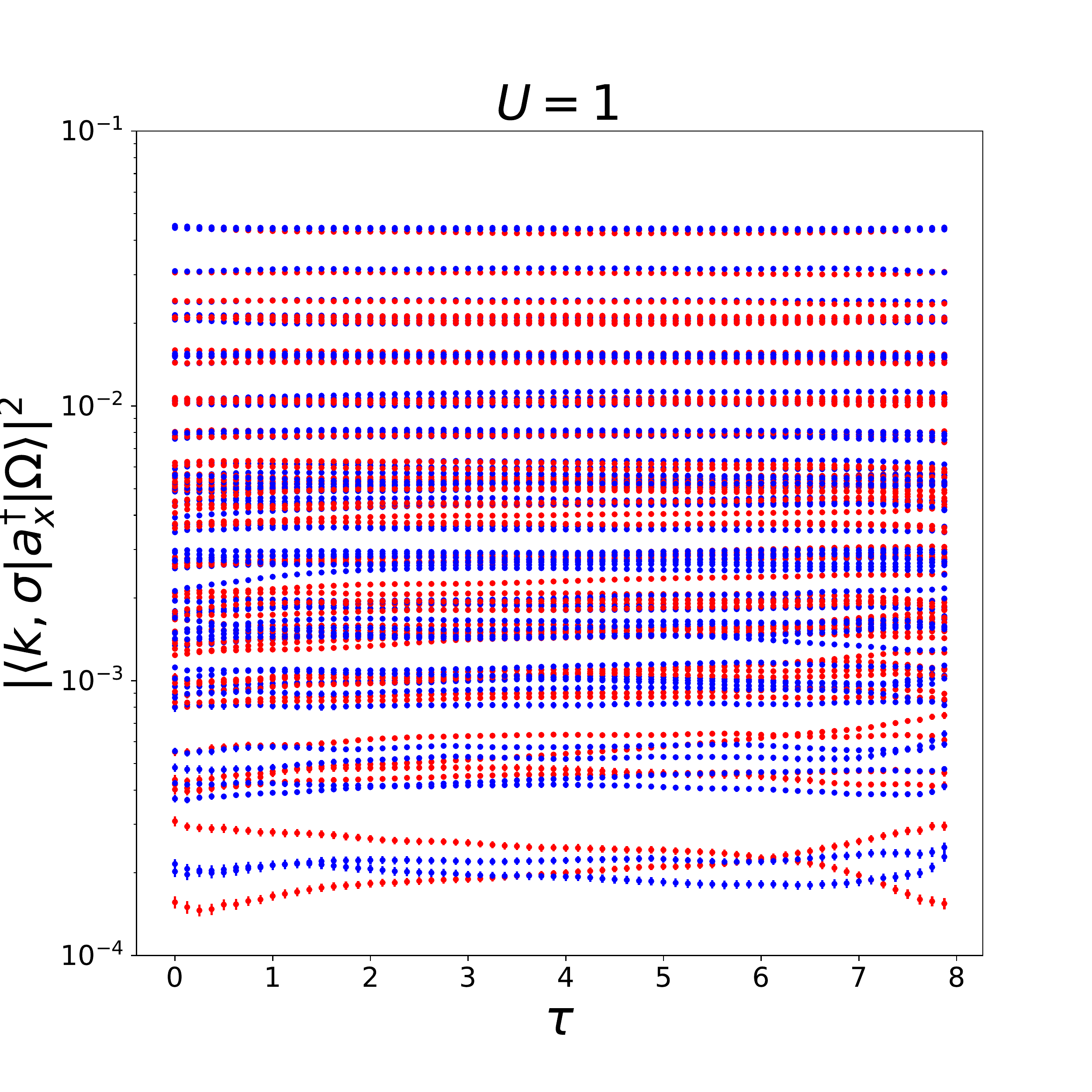}\includegraphics[width=.5\textwidth]{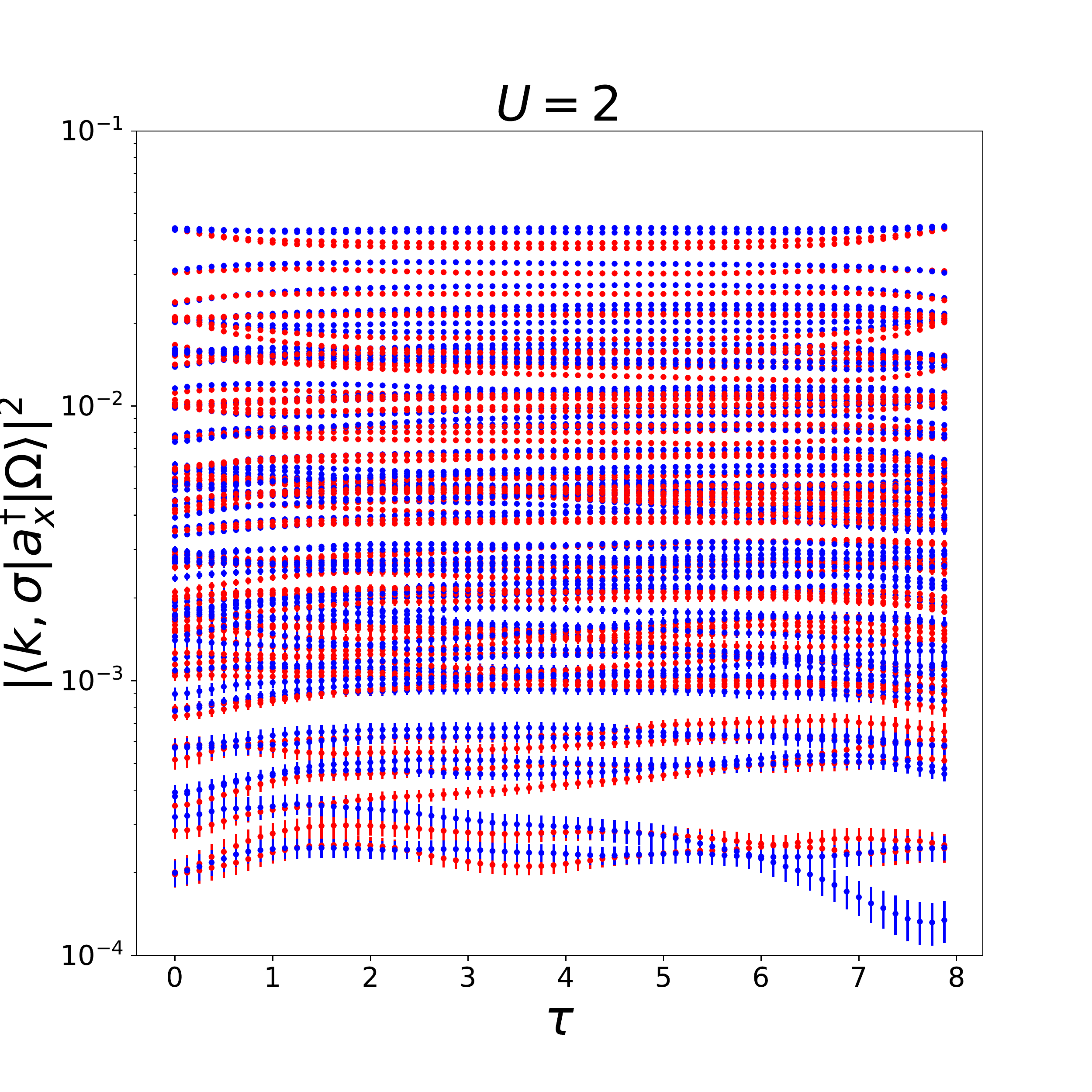}\\
    \includegraphics[width=.5\textwidth]{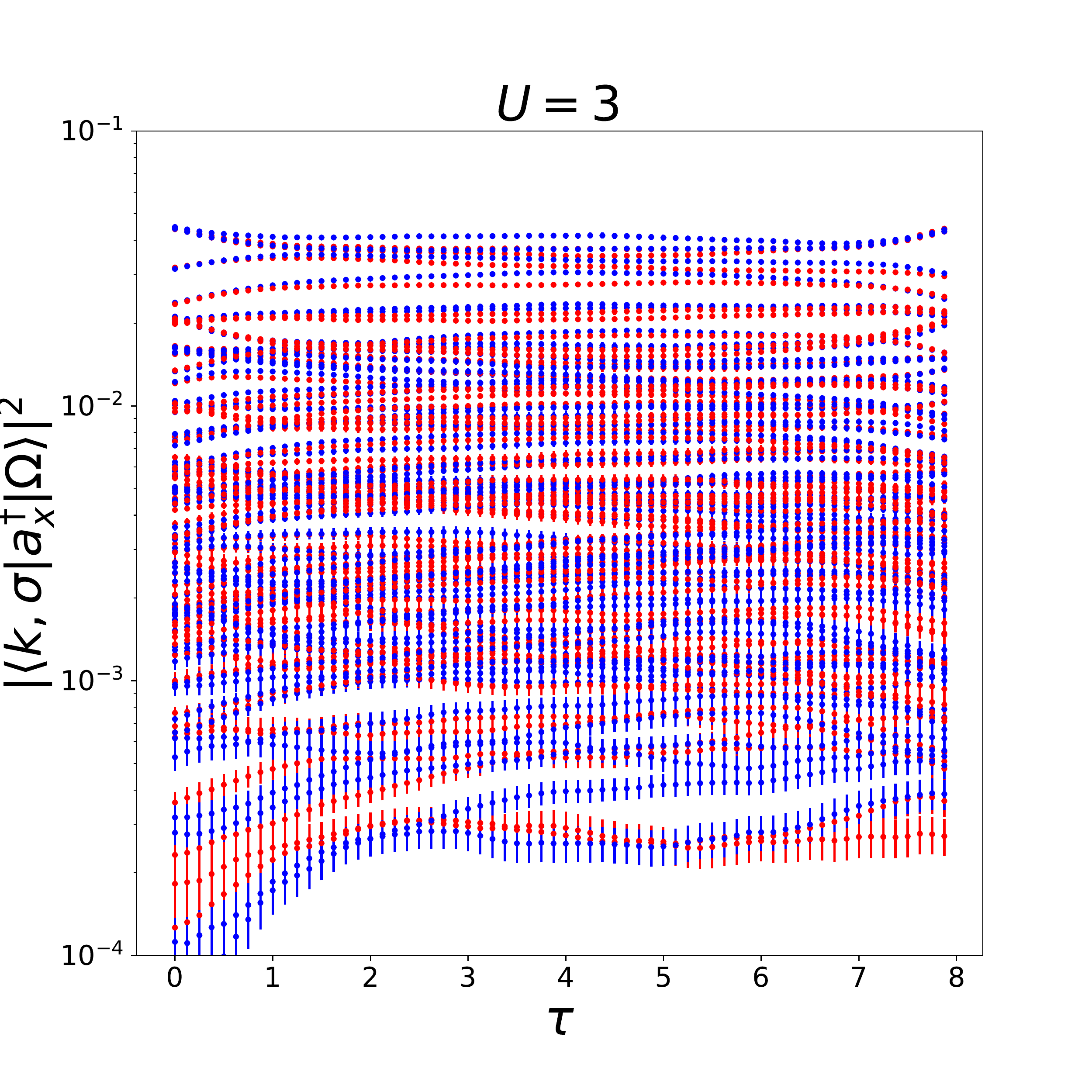}\includegraphics[width=.5\textwidth]{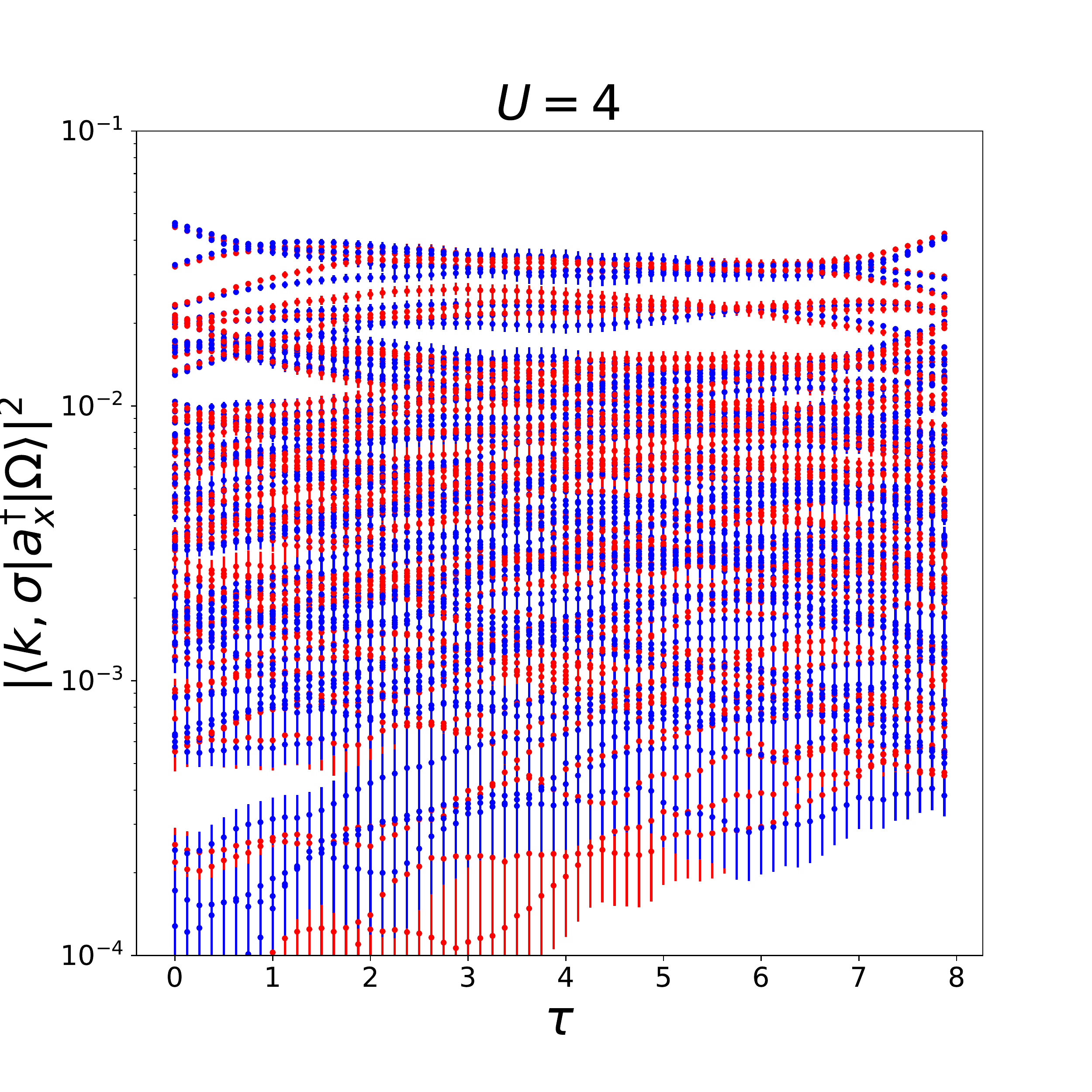}
    \caption{Same as is in Fig.~\ref{fig:non interacting amplitudes}, but now with non-zero values $U$.  The non-interacting amplitudes at $U=0$ are shown in Fig.~\ref{fig:non interacting amplitudes}.}
    \label{fig:amplitudes}
\end{figure}

The amplitude we are interested in is $\langle \Omega|a_x|\Omega+k\rangle$.  Note that in the non-interacting limit we have that $\langle \Omega+k|a^\dag_k|\Omega\rangle=1$ and the amplitude is, up to an overall phase, equivalent to $C^\kappa_i$ in Eq.~\eqref{eqn:ak}.  With interactions, unfortunately, we cannot extract this amplitude because it is multiplied by the factor $\langle \Omega+k|a^\dag_k|\Omega\rangle e^{-(E_{\Omega+k}-E_\Omega)t}$ which we do not \emph{a priori} know.  However, note that this factor is \emph{independent} of the site $x$ and carries the same time dependence for all spatial correlators.  Furthermore, we are interested in the densities, $\rho_k(x)=|\langle \Omega|a_x|\Omega+k\rangle|^2$ which should be normalized over the lattice unit cell, $\sum_x\rho_k(x)=1$.  With these properties in mind, we instead analyze the following expression,
\begin{equation}\label{eqn:density definition}
    \rho_k(x,\tau)\equiv \frac{|C_k(x,\tau)|^2}{\sum_y |C_k(y,\tau)|^2}\ .
\end{equation}
Because of the independence of the unknown factor on spatial site $x$ and its identical time dependence for each spatial site, this factor cancels in this ratio.  The resulting term is automatically normalized over all lattice sites and thus represents the density at each site $x$.  In the non-interacting limit, the cancellation of the unknown factor occurs exactly for all $\tau$, and so Eq.~\eqref{eqn:density definition} has no dependence on $\tau$.  We have verified that it produces the exact wavefunction densities, as shown in Fig.~\ref{fig:non interacting amplitudes}.  For $U\ne0$, the cancellation of the unknown factor occurs only in the scaling region given in Eq.~\eqref{eqn:scaling behavior}, and so we extract the densities in the region where $\rho_k(x,t)$ exhibits little to no time dependence and is thus relatively flat.  Figure~\ref{fig:amplitudes} shows examples of the $\rho_k(x,t)$ for different values of $U$ including the non-interacting case.  In all cases we extract the density in a region centered around $t=\beta/2$.


\begin{thebibliography}{16}%
\makeatletter
\providecommand \@ifxundefined [1]{%
 \@ifx{#1\undefined}
}%
\providecommand \@ifnum [1]{%
 \ifnum #1\expandafter \@firstoftwo
 \else \expandafter \@secondoftwo
 \fi
}%
\providecommand \@ifx [1]{%
 \ifx #1\expandafter \@firstoftwo
 \else \expandafter \@secondoftwo
 \fi
}%
\providecommand \natexlab [1]{#1}%
\providecommand \enquote  [1]{``#1''}%
\providecommand \bibnamefont  [1]{#1}%
\providecommand \bibfnamefont [1]{#1}%
\providecommand \citenamefont [1]{#1}%
\providecommand \href@noop [0]{\@secondoftwo}%
\providecommand \href [0]{\begingroup \@sanitize@url \@href}%
\providecommand \@href[1]{\@@startlink{#1}\@@href}%
\providecommand \@@href[1]{\endgroup#1\@@endlink}%
\providecommand \@sanitize@url [0]{\catcode `\\12\catcode `\$12\catcode
  `\&12\catcode `\#12\catcode `\^12\catcode `\_12\catcode `\%12\relax}%
\providecommand \@@startlink[1]{}%
\providecommand \@@endlink[0]{}%
\providecommand \url  [0]{\begingroup\@sanitize@url \@url }%
\providecommand \@url [1]{\endgroup\@href {#1}{\urlprefix }}%
\providecommand \urlprefix  [0]{URL }%
\providecommand \Eprint [0]{\href }%
\providecommand \doibase [0]{https://doi.org/}%
\providecommand \selectlanguage [0]{\@gobble}%
\providecommand \bibinfo  [0]{\@secondoftwo}%
\providecommand \bibfield  [0]{\@secondoftwo}%
\providecommand \translation [1]{[#1]}%
\providecommand \BibitemOpen [0]{}%
\providecommand \bibitemStop [0]{}%
\providecommand \bibitemNoStop [0]{.\EOS\space}%
\providecommand \EOS [0]{\spacefactor3000\relax}%
\providecommand \BibitemShut  [1]{\csname bibitem#1\endcsname}%
\let\auto@bib@innerbib\@empty
%</preamble>
\bibitem [{\citenamefont {Cao}\ \emph {et~al.}(2017)\citenamefont {Cao},
  \citenamefont {Zhao},\ and\ \citenamefont {Louie}}]{PhysRevLett.119.076401}%
  \BibitemOpen
  \bibfield  {author} {\bibinfo {author} {\bibfnamefont {T.}~\bibnamefont
  {Cao}}, \bibinfo {author} {\bibfnamefont {F.}~\bibnamefont {Zhao}},\ and\
  \bibinfo {author} {\bibfnamefont {S.~G.}\ \bibnamefont {Louie}},\ }\bibfield
  {title} {\bibinfo {title} {Topological phases in graphene nanoribbons:
  Junction states, spin centers, and quantum spin chains},\ }\href
  {https://doi.org/10.1103/PhysRevLett.119.076401} {\bibfield  {journal}
  {\bibinfo  {journal} {Phys. Rev. Lett.}\ }\textbf {\bibinfo {volume} {119}},\
  \bibinfo {pages} {076401} (\bibinfo {year} {2017})}\BibitemShut {NoStop}%
\bibitem [{\citenamefont {Rizzo}\ \emph {et~al.}(2018)\citenamefont {Rizzo},
  \citenamefont {Veber}, \citenamefont {Cao}, \citenamefont {Bronner},
  \citenamefont {Chen}, \citenamefont {Zhao}, \citenamefont {Rodriguez},
  \citenamefont {Louie}, \citenamefont {Crommie},\ and\ \citenamefont
  {Fischer}}]{rizzo18}%
  \BibitemOpen
  \bibfield  {author} {\bibinfo {author} {\bibfnamefont {D.~J.}\ \bibnamefont
  {Rizzo}}, \bibinfo {author} {\bibfnamefont {G.}~\bibnamefont {Veber}},
  \bibinfo {author} {\bibfnamefont {T.}~\bibnamefont {Cao}}, \bibinfo {author}
  {\bibfnamefont {C.}~\bibnamefont {Bronner}}, \bibinfo {author} {\bibfnamefont
  {T.}~\bibnamefont {Chen}}, \bibinfo {author} {\bibfnamefont {F.}~\bibnamefont
  {Zhao}}, \bibinfo {author} {\bibfnamefont {H.}~\bibnamefont {Rodriguez}},
  \bibinfo {author} {\bibfnamefont {S.~G.}\ \bibnamefont {Louie}}, \bibinfo
  {author} {\bibfnamefont {M.~F.}\ \bibnamefont {Crommie}},\ and\ \bibinfo
  {author} {\bibfnamefont {F.~R.}\ \bibnamefont {Fischer}},\ }\bibfield
  {title} {\bibinfo {title} {Topological band engineering of graphene
  nanoribbons},\ }\href {https://doi.org/10.1038/s41586-018-0376-8} {\bibfield
  {journal} {\bibinfo  {journal} {Nature}\ }\textbf {\bibinfo {volume} {560}},\
  \bibinfo {pages} {204} (\bibinfo {year} {2018})}\BibitemShut {NoStop}%
\bibitem [{\citenamefont {Gr{\"o}ning}\ \emph {et~al.}(2018)\citenamefont
  {Gr{\"o}ning}, \citenamefont {Wang}, \citenamefont {Yao}, \citenamefont
  {Pignedoli}, \citenamefont {Borin~Barin}, \citenamefont {Daniels},
  \citenamefont {Cupo}, \citenamefont {Meunier}, \citenamefont {Feng},
  \citenamefont {Narita}, \citenamefont {M{\"u}llen}, \citenamefont
  {Ruffieux},\ and\ \citenamefont {Fasel}}]{groening2018}%
  \BibitemOpen
  \bibfield  {author} {\bibinfo {author} {\bibfnamefont {O.}~\bibnamefont
  {Gr{\"o}ning}}, \bibinfo {author} {\bibfnamefont {S.}~\bibnamefont {Wang}},
  \bibinfo {author} {\bibfnamefont {X.}~\bibnamefont {Yao}}, \bibinfo {author}
  {\bibfnamefont {C.~A.}\ \bibnamefont {Pignedoli}}, \bibinfo {author}
  {\bibfnamefont {G.}~\bibnamefont {Borin~Barin}}, \bibinfo {author}
  {\bibfnamefont {C.}~\bibnamefont {Daniels}}, \bibinfo {author} {\bibfnamefont
  {A.}~\bibnamefont {Cupo}}, \bibinfo {author} {\bibfnamefont {V.}~\bibnamefont
  {Meunier}}, \bibinfo {author} {\bibfnamefont {X.}~\bibnamefont {Feng}},
  \bibinfo {author} {\bibfnamefont {A.}~\bibnamefont {Narita}}, \bibinfo
  {author} {\bibfnamefont {K.}~\bibnamefont {M{\"u}llen}}, \bibinfo {author}
  {\bibfnamefont {P.}~\bibnamefont {Ruffieux}},\ and\ \bibinfo {author}
  {\bibfnamefont {R.}~\bibnamefont {Fasel}},\ }\bibfield  {title} {\bibinfo
  {title} {Engineering of robust topological quantum phases in graphene
  nanoribbons},\ }\href {https://doi.org/10.1038/s41586-018-0375-9} {\bibfield
  {journal} {\bibinfo  {journal} {Nature}\ }\textbf {\bibinfo {volume} {560}},\
  \bibinfo {pages} {209} (\bibinfo {year} {2018})}\BibitemShut {NoStop}%
\bibitem [{\citenamefont {Rizzo}\ \emph {et~al.}(2021)\citenamefont {Rizzo},
  \citenamefont {Jiang}, \citenamefont {Joshi}, \citenamefont {Veber},
  \citenamefont {Bronner}, \citenamefont {Durr}, \citenamefont {Jacobse},
  \citenamefont {Cao}, \citenamefont {Kalayjian}, \citenamefont {Rodriguez},
  \citenamefont {Butler}, \citenamefont {Chen}, \citenamefont {Louie},
  \citenamefont {Fischer},\ and\ \citenamefont
  {Crommie}}]{doi:10.1021/acsnano.1c09503}%
  \BibitemOpen
  \bibfield  {author} {\bibinfo {author} {\bibfnamefont {D.~J.}\ \bibnamefont
  {Rizzo}}, \bibinfo {author} {\bibfnamefont {J.}~\bibnamefont {Jiang}},
  \bibinfo {author} {\bibfnamefont {D.}~\bibnamefont {Joshi}}, \bibinfo
  {author} {\bibfnamefont {G.}~\bibnamefont {Veber}}, \bibinfo {author}
  {\bibfnamefont {C.}~\bibnamefont {Bronner}}, \bibinfo {author} {\bibfnamefont
  {R.~A.}\ \bibnamefont {Durr}}, \bibinfo {author} {\bibfnamefont {P.~H.}\
  \bibnamefont {Jacobse}}, \bibinfo {author} {\bibfnamefont {T.}~\bibnamefont
  {Cao}}, \bibinfo {author} {\bibfnamefont {A.}~\bibnamefont {Kalayjian}},
  \bibinfo {author} {\bibfnamefont {H.}~\bibnamefont {Rodriguez}}, \bibinfo
  {author} {\bibfnamefont {P.}~\bibnamefont {Butler}}, \bibinfo {author}
  {\bibfnamefont {T.}~\bibnamefont {Chen}}, \bibinfo {author} {\bibfnamefont
  {S.~G.}\ \bibnamefont {Louie}}, \bibinfo {author} {\bibfnamefont {F.~R.}\
  \bibnamefont {Fischer}},\ and\ \bibinfo {author} {\bibfnamefont {M.~F.}\
  \bibnamefont {Crommie}},\ }\bibfield  {title} {\bibinfo {title} {Rationally
  designed topological quantum dots in bottom-up graphene nanoribbons},\ }\href
  {https://doi.org/10.1021/acsnano.1c09503} {\bibfield  {journal} {\bibinfo
  {journal} {ACS Nano}\ }\textbf {\bibinfo {volume} {15}},\ \bibinfo {pages}
  {20633} (\bibinfo {year} {2021})},\ \bibinfo {note} {pMID: 34842409},\
  \Eprint {https://arxiv.org/abs/https://doi.org/10.1021/acsnano.1c09503}
  {https://doi.org/10.1021/acsnano.1c09503} \BibitemShut {NoStop}%
\bibitem [{\citenamefont {Yang}\ \emph {et~al.}(2020)\citenamefont {Yang},
  \citenamefont {Cha}, \citenamefont {Lee},\ and\ \citenamefont
  {Kim}}]{Yang:2020lal}%
  \BibitemOpen
  \bibfield  {author} {\bibinfo {author} {\bibfnamefont {S.~R.~E.}\
  \bibnamefont {Yang}}, \bibinfo {author} {\bibfnamefont {M.-C.}\ \bibnamefont
  {Cha}}, \bibinfo {author} {\bibfnamefont {H.~J.}\ \bibnamefont {Lee}},\ and\
  \bibinfo {author} {\bibfnamefont {Y.~H.}\ \bibnamefont {Kim}},\ }\bibfield
  {title} {\bibinfo {title} {{Topologically ordered zigzag nanoribbon: $e/2$
  fractional edge charge, spin-charge separation, and ground state
  degeneracy}},\ }\href {https://doi.org/10.1103/PhysRevResearch.2.033109}
  {\bibfield  {journal} {\bibinfo  {journal} {Phys. Rev. Res.}\ }\textbf
  {\bibinfo {volume} {2}},\ \bibinfo {pages} {033109} (\bibinfo {year}
  {2020})},\ \Eprint {https://arxiv.org/abs/2004.14125} {arXiv:2004.14125
  [cond-mat.str-el]} \BibitemShut {NoStop}%
\bibitem [{\citenamefont {Ezawa}(2018)}]{Ezawa2018}%
  \BibitemOpen
  \bibfield  {author} {\bibinfo {author} {\bibfnamefont {M.}~\bibnamefont
  {Ezawa}},\ }\bibfield  {title} {\bibinfo {title} {Exact solutions for
  two-dimensional topological superconductors: Hubbard interaction induced
  spontaneous symmetry breaking},\ }\href
  {https://doi.org/10.1103/PhysRevB.97.241113} {\bibfield  {journal} {\bibinfo
  {journal} {Phys. Rev. B}\ }\textbf {\bibinfo {volume} {97}},\ \bibinfo
  {pages} {241113} (\bibinfo {year} {2018})}\BibitemShut {NoStop}%
\bibitem [{\citenamefont {Miao}\ \emph {et~al.}(2019)\citenamefont {Miao},
  \citenamefont {Xu}, \citenamefont {Zhang},\ and\ \citenamefont
  {Zhang}}]{Miao:2019tng}%
  \BibitemOpen
  \bibfield  {author} {\bibinfo {author} {\bibfnamefont {J.-J.}\ \bibnamefont
  {Miao}}, \bibinfo {author} {\bibfnamefont {D.-H.}\ \bibnamefont {Xu}},
  \bibinfo {author} {\bibfnamefont {L.}~\bibnamefont {Zhang}},\ and\ \bibinfo
  {author} {\bibfnamefont {F.-C.}\ \bibnamefont {Zhang}},\ }\bibfield  {title}
  {\bibinfo {title} {{Exact solution to the Haldane-BCS-Hubbard model along the
  symmetric lines: Interaction-induced topological phase transition}},\ }\href
  {https://doi.org/10.1103/PhysRevB.99.245154} {\bibfield  {journal} {\bibinfo
  {journal} {Phys. Rev. B}\ }\textbf {\bibinfo {volume} {99}},\ \bibinfo
  {pages} {245154} (\bibinfo {year} {2019})},\ \Eprint
  {https://arxiv.org/abs/1903.06101} {arXiv:1903.06101 [cond-mat.str-el]}
  \BibitemShut {NoStop}%
\bibitem [{\citenamefont {Kaplan}(1992)}]{Kaplan:1992bt}%
  \BibitemOpen
  \bibfield  {author} {\bibinfo {author} {\bibfnamefont {D.~B.}\ \bibnamefont
  {Kaplan}},\ }\bibfield  {title} {\bibinfo {title} {{A Method for simulating
  chiral fermions on the lattice}},\ }\href
  {https://doi.org/10.1016/0370-2693(92)91112-M} {\bibfield  {journal}
  {\bibinfo  {journal} {Phys. Lett. B}\ }\textbf {\bibinfo {volume} {288}},\
  \bibinfo {pages} {342} (\bibinfo {year} {1992})},\ \Eprint
  {https://arxiv.org/abs/hep-lat/9206013} {arXiv:hep-lat/9206013} \BibitemShut
  {NoStop}%
\bibitem [{\citenamefont {Shamir}(1993)}]{Shamir:1993zy}%
  \BibitemOpen
  \bibfield  {author} {\bibinfo {author} {\bibfnamefont {Y.}~\bibnamefont
  {Shamir}},\ }\bibfield  {title} {\bibinfo {title} {{Chiral fermions from
  lattice boundaries}},\ }\href {https://doi.org/10.1016/0550-3213(93)90162-I}
  {\bibfield  {journal} {\bibinfo  {journal} {Nucl. Phys. B}\ }\textbf
  {\bibinfo {volume} {406}},\ \bibinfo {pages} {90} (\bibinfo {year} {1993})},\
  \Eprint {https://arxiv.org/abs/hep-lat/9303005} {arXiv:hep-lat/9303005}
  \BibitemShut {NoStop}%
\bibitem [{\citenamefont {Rhim}\ \emph {et~al.}(2017)\citenamefont {Rhim},
  \citenamefont {Behrends},\ and\ \citenamefont
  {Bardarson}}]{PhysRevB.95.035421}%
  \BibitemOpen
  \bibfield  {author} {\bibinfo {author} {\bibfnamefont {J.-W.}\ \bibnamefont
  {Rhim}}, \bibinfo {author} {\bibfnamefont {J.}~\bibnamefont {Behrends}},\
  and\ \bibinfo {author} {\bibfnamefont {J.~H.}\ \bibnamefont {Bardarson}},\
  }\bibfield  {title} {\bibinfo {title} {Bulk-boundary correspondence from the
  intercellular zak phase},\ }\href
  {https://doi.org/10.1103/PhysRevB.95.035421} {\bibfield  {journal} {\bibinfo
  {journal} {Phys. Rev. B}\ }\textbf {\bibinfo {volume} {95}},\ \bibinfo
  {pages} {035421} (\bibinfo {year} {2017})}\BibitemShut {NoStop}%
\bibitem [{\citenamefont {Luu}\ and\ \citenamefont
  {L\"ahde}(2016)}]{Luu:2015gpl}%
  \BibitemOpen
  \bibfield  {author} {\bibinfo {author} {\bibfnamefont {T.}~\bibnamefont
  {Luu}}\ and\ \bibinfo {author} {\bibfnamefont {T.~A.}\ \bibnamefont
  {L\"ahde}},\ }\bibfield  {title} {\bibinfo {title} {{Quantum Monte Carlo
  Calculations for Carbon Nanotubes}},\ }\href
  {https://doi.org/10.1103/PhysRevB.93.155106} {\bibfield  {journal} {\bibinfo
  {journal} {Phys. Rev. B}\ }\textbf {\bibinfo {volume} {93}},\ \bibinfo
  {pages} {155106} (\bibinfo {year} {2016})},\ \Eprint
  {https://arxiv.org/abs/1511.04918} {arXiv:1511.04918 [cond-mat.str-el]}
  \BibitemShut {NoStop}%
\bibitem [{\citenamefont {Ostmeyer}\ \emph {et~al.}(2020)\citenamefont
  {Ostmeyer}, \citenamefont {Berkowitz}, \citenamefont {Krieg}, \citenamefont
  {L\"ahde}, \citenamefont {Luu},\ and\ \citenamefont
  {Urbach}}]{Ostmeyer:2020uov}%
  \BibitemOpen
  \bibfield  {author} {\bibinfo {author} {\bibfnamefont {J.}~\bibnamefont
  {Ostmeyer}}, \bibinfo {author} {\bibfnamefont {E.}~\bibnamefont {Berkowitz}},
  \bibinfo {author} {\bibfnamefont {S.}~\bibnamefont {Krieg}}, \bibinfo
  {author} {\bibfnamefont {T.~A.}\ \bibnamefont {L\"ahde}}, \bibinfo {author}
  {\bibfnamefont {T.}~\bibnamefont {Luu}},\ and\ \bibinfo {author}
  {\bibfnamefont {C.}~\bibnamefont {Urbach}},\ }\bibfield  {title} {\bibinfo
  {title} {{Semimetal\textendash{}Mott insulator quantum phase transition of
  the Hubbard model on the honeycomb lattice}},\ }\href
  {https://doi.org/10.1103/PhysRevB.102.245105} {\bibfield  {journal} {\bibinfo
   {journal} {Phys. Rev. B}\ }\textbf {\bibinfo {volume} {102}},\ \bibinfo
  {pages} {245105} (\bibinfo {year} {2020})},\ \Eprint
  {https://arxiv.org/abs/2005.11112} {arXiv:2005.11112 [cond-mat.str-el]}
  \BibitemShut {NoStop}%
\bibitem [{\citenamefont {Ostmeyer}\ \emph {et~al.}(2021)\citenamefont
  {Ostmeyer}, \citenamefont {Berkowitz}, \citenamefont {Krieg}, \citenamefont
  {L\"ahde}, \citenamefont {Luu},\ and\ \citenamefont
  {Urbach}}]{Ostmeyer:2021efs}%
  \BibitemOpen
  \bibfield  {author} {\bibinfo {author} {\bibfnamefont {J.}~\bibnamefont
  {Ostmeyer}}, \bibinfo {author} {\bibfnamefont {E.}~\bibnamefont {Berkowitz}},
  \bibinfo {author} {\bibfnamefont {S.}~\bibnamefont {Krieg}}, \bibinfo
  {author} {\bibfnamefont {T.~A.}\ \bibnamefont {L\"ahde}}, \bibinfo {author}
  {\bibfnamefont {T.}~\bibnamefont {Luu}},\ and\ \bibinfo {author}
  {\bibfnamefont {C.}~\bibnamefont {Urbach}},\ }\bibfield  {title} {\bibinfo
  {title} {{Antiferromagnetic character of the quantum phase transition in the
  Hubbard model on the honeycomb lattice}},\ }\href
  {https://doi.org/10.1103/PhysRevB.104.155142} {\bibfield  {journal} {\bibinfo
   {journal} {Phys. Rev. B}\ }\textbf {\bibinfo {volume} {104}},\ \bibinfo
  {pages} {155142} (\bibinfo {year} {2021})},\ \Eprint
  {https://arxiv.org/abs/2105.06936} {arXiv:2105.06936 [cond-mat.str-el]}
  \BibitemShut {NoStop}%
\bibitem [{\citenamefont {Bogolyubov}(1958)}]{Bogolyubov:1958km}%
  \BibitemOpen
  \bibfield  {author} {\bibinfo {author} {\bibfnamefont {N.~N.}\ \bibnamefont
  {Bogolyubov}},\ }\bibfield  {title} {\bibinfo {title} {{On a New method in
  the theory of superconductivity}},\ }\href
  {https://doi.org/10.1007/BF02745585} {\bibfield  {journal} {\bibinfo
  {journal} {Nuovo Cim.}\ }\textbf {\bibinfo {volume} {7}},\ \bibinfo {pages}
  {794} (\bibinfo {year} {1958})}\BibitemShut {NoStop}%
\bibitem [{\citenamefont {Valatin}(1958)}]{Valatin:1958ja}%
  \BibitemOpen
  \bibfield  {author} {\bibinfo {author} {\bibfnamefont {J.~G.}\ \bibnamefont
  {Valatin}},\ }\bibfield  {title} {\bibinfo {title} {{Comments on the theory
  of superconductivity}},\ }\href {https://doi.org/10.1007/BF02745589}
  {\bibfield  {journal} {\bibinfo  {journal} {Nuovo Cim.}\ }\textbf {\bibinfo
  {volume} {7}},\ \bibinfo {pages} {843} (\bibinfo {year} {1958})}\BibitemShut
  {NoStop}%
\bibitem [{\citenamefont {Nielsen}\ and\ \citenamefont
  {Ninomiya}(1981)}]{Nielsen:1981hk}%
  \BibitemOpen
  \bibfield  {author} {\bibinfo {author} {\bibfnamefont {H.~B.}\ \bibnamefont
  {Nielsen}}\ and\ \bibinfo {author} {\bibfnamefont {M.}~\bibnamefont
  {Ninomiya}},\ }\bibfield  {title} {\bibinfo {title} {{No Go Theorem for
  Regularizing Chiral Fermions}},\ }\href
  {https://doi.org/10.1016/0370-2693(81)91026-1} {\bibfield  {journal}
  {\bibinfo  {journal} {Phys. Lett. B}\ }\textbf {\bibinfo {volume} {105}},\
  \bibinfo {pages} {219} (\bibinfo {year} {1981})}\BibitemShut {NoStop}%
\end{thebibliography}
\end{document}